\documentclass[fleqn,usenatbib]{mnras}
\usepackage{graphicx}
\usepackage{dcolumn}
\usepackage{bm}
\usepackage{float}
\usepackage{ amssymb }

\title[ALP hint from z dependence of blazar spectra]
{Hint at an axion-like particle from the redshift dependence of blazar spectra}

\author[G. Galanti et al.]{
G. Galanti,$^{1}$\thanks{E-mail: gam.galanti@gmail.com (GG)}
M. Roncadelli,$^{2,3}$
A. De Angelis$^{4,5}$
and G. F. Bignami$^{6}$\thanks{Deceased}
\\
$^{1}$INAF, Osservatorio Astronomico di Brera, Via E. Bianchi 46, I -- 23807 Merate, Italy\\
$^{2}$INFN, Sezione di Pavia, Via A. Bassi 6, I -- 27100 Pavia, Italy\\
$^{3}$INAF, IASF Milano, Via A. Corti 12, I -- 20133 Milano, Italy\\
$^{4}$Dipartimento di Fisica e Astronomia G. Galilei, Via F. Marzolo 8, I -- 35131 Padova, Italy\\
$^{5}$INFN, Sezione di Padova, Via F. Marzolo 8, I -- 35131 Padova, Italy\\
$^{6}$Accademia dei Lincei, Via della Lungara 10, I -- 00165 Roma, Italy
}

\date{Accepted XXX. Received YYY; in original form ZZZ}

\pubyear{2019}

\begin{document}
\label{firstpage}
\pagerange{\pageref{firstpage}--\pageref{lastpage}}
\maketitle

\begin{abstract}
We consider the largest observed sample including all intermediate-frequency peaked (IBL) and high-frequency peaked (HBL) flaring blazars above 100 GeV up to redshift $z = 0.6$. We show that the best-fit regression line of the emitted spectral indices $\Gamma_{\rm em} (z)$ is a concave parabola decreasing as $z$ increases, thereby implying a {\it statistical correlation} between the $\{\Gamma_{\rm em} (z) \}$ distribution and $z$. This result contradicts our expectation that such a distribution should be $z$-{\it independent}. We argue that the above correlation does not arise from any selection bias. We show that our expectation {\it naturally emerges} provided that axion-like particles (ALPs) are put into the game.  Moreover, ALPs can also explain why flat spectrum radio quasars emit up to 400 GeV, in sharp contradiction with conventional physics. So, the combination of the two very different but consistent results -- taken at face value -- leads to a hint at an ALP with mass $m = {\cal O} (10^{-10} \, {\rm eV})$ and two-photon coupling in the range $2.94 \times 10^{- 12} \, {\rm GeV}^{- 1} < g_{a \gamma \gamma} < 0.66 \times 10^{- 10} \, {\rm GeV}^{- 1}$. As a bonus, the Universe would become  considerably more transparent above energies $E \gtrsim 1 \, {\rm TeV}$ than dictated by conventional physics. Our prediction can be checked not only by the new generation of observatories like CTA, HAWC, GAMMA-400, LHAASO, TAIGA-HiSCORE and HERD, but also thanks to the planned laboratory experiments ALPS II (upgraded), STAX, IAXO and with other techniques now being developed by Avignone and collaborators.

\center{\it We wish to dedicate the present work to the memory of our dear friend

Nanni Bignami.}
\end{abstract}




\section{INTRODUCTION}

Thanks to the observations carried out with the Imaging Atmospheric Che\-ren\-kov Telescopes (IACTs) H.E.S.S. (High Energy Stereoscopic System)\footnote{https://www.mpi-hd.mpg.de/hfm/HESS}, MAGIC (Major Atmospheric Gamma Imaging Cherenkov Telescopes)\footnote{https://magic.mpp.mpg.de/} and VERITAS (Very Energetic Radiation Imaging Telescope Array System)\footnote{http://veritas.sao.arizona.edu/}, according to the Tevcat catalog more than 50 blazars have been detected so far in the very-high-energy (VHE) band ($100 \, {\rm GeV} - 100 \, {\rm TeV}$)\footnote{http://Tevcat.uchicago.edu/}. We would like to stress  that even though a single collaboration (H.E.S.S., MAGIC, VERITAS) has only a partial sky coverage, H.E.S.S. is located in Namibia, MAGIC in one of the Canary Islands and VERITAS in Arizona. Because roughly one-third of the considered sources are observed by each collaboration, their combination yields a quasi complete sky coverage with the only exception of the North and South Poles. 

Unfortunately, not all these sources are suitable for the study carried out in the present paper. The reason is four-fold (henceforth we use the terms BL Lac, blazar and source interchangeably). 

\begin{enumerate} 

\item We restrict our attention to {\it flaring} blazars, namely those showing episodes of time variability with their luminosity increasing by more than a factor of two, ranging from a few hours to a few days: the reason is both their enhanced luminosity -- and so their detectability~\citep{flare1,flare2} -- and our desire to deal with a homogeneous sample. 

\item In view of our analysis, we need to know the redshift, the observed spectrum and the energy range in which every blazar is observed. This information is available only for a subset of the observed flaring sources. 

\item We want to avoid cosmological evolutionary effects inside blazars: this requirement leads us to consider only sources up to $z = 0.6$.

\item Finally, we want to work with a source sample which is as homogeneous as possible. For this reason we consider only intermediate-frequency peaked (IBL) and high-frequency peaked (HBL) flaring BL Lacs with observed energy $E_0 \gtrsim 100 \, {\rm GeV}$.

\end{enumerate} 

We are consequently left with a sample ${\cal S}$ of 39 flaring VHE BL Lacs, which are listed in Table I of the Supplementary Material (SM). 

Given the preliminary character of the present investigation, all observed spectra of the VHE blazars in ${\cal S}$ are fitted by a single power-law -- neglecting a possible small curvature of some spectra in their lowest energy part -- and so they have the form $\Phi_{\rm obs}(E_0, z) = \hat{K}_{\rm obs} (z) \, E_0^{- \Gamma_{\rm obs}(z)}$, where $E_0$ is the observed energy, while $\hat{K}_{\rm obs} (z)$ and $\Gamma_{\rm obs} (z)$ denote the normalization constant and the observed slope, respectively, for a source at redshift $z$. 

As a matter of fact, the observational results do not provide any {\it direct} information about the {\it intrinsic} properties of the sources, since the VHE gamma-ray data strongly depend on the nature of photon propagation. Indeed, according to conventional physics the blazar spectra in the VHE band are strongly affected by the presence of the Extragalactic Background Light (EBL), namely the infrared/optical/ultraviolet background photons emitted by all stars since their birth (for a review, see~\citealt{ebl}). But it should be kept in mind that if some yet-to-be-discovered new physics changes the photon propagation, then some intrinsic source properties that are currently believed to be true may actually turn out to be incorrect.

The conventional view about the observations of blazars with the IACTs is that the VHE photons impinging on the atmosphere are produced by the blazars themselves through {\it photon emission models}, which can involve either a leptonic~\citep{ssc1,ssc2} or a hadronic mechanism~\citep{mannheim1,mannheim2}. In fact, the above-mentioned three IACT collaborations systematically analyze their data using both models. The alternative possibility of {\it proton emission models}~\citep{esseykusenko2010} cannot explain flaring blazars at the presently probed energies. Indeed, an observed blazars variability shorter than $0.1 \, {\rm yr}$ can be explained by the proton emission model only for $z > 0.20$ at $E_0 > 1 \, {\rm  TeV}$~\citep{prosekinesseykusenkoaharonian2012}. Hence,  throughout this paper we will focus our attention on photon emission models. 

Before proceeding further, a warning to the reader is of paramount important. Whenever we talk about {\it emission} we mean -- according to the standard practice in this field -- `EBL-deabsorbed', the reason being that the EBL-absorption changes the shape of the spectra, whereas the Malmquist bias does not. Because we are concerned throughout this paper only with the shape-dependence of BL Lac spectra on $z$, the Malmquist bias will be neglected altogether. We also recall that the observed spectrum is $\Phi_{\rm obs}(E_0, z) \equiv d N_{\rm obs}/(d A \, d t \, d E_0)$, where $N_{\rm obs}$ is the photon number count and $A$ is the detector area, and likewise the emitted flux is $\Phi_{\rm em} (E) \equiv d N_{\rm em}/(d A \, d t \, d E)$, where $N_{\rm em}$ is the number of emitted photons and $A$ is the area of the emitting region. Now, both photon emission models predict emitted spectra which -- to a good approximation -- have a single power-law behavior $\Phi_{\rm em} (E) = \hat{K}_{\rm em} \, E^{- \Gamma_{\rm em}}$ for all IBL and HBL VHE blazars in ${\cal S}$, where $\hat{K}_{\rm em}$ is the normalization constant and $\Gamma_{\rm em}$ is the emitted slope (the specific values of $\hat{K}_{\rm em}$ and $\Gamma_{\rm em}$ vary from source to source). 

The relationship between $\Phi_{\rm obs}(E_0, z)$ and $\Phi_{\rm em}(E)$ can be generally expressed as 
\begin{equation}
\Phi_{\rm obs}(E_0, z) = P_{\gamma \to \gamma} (E_0, z) \, \Phi_{\rm em} \bigl(E_0 (1 + z) \bigr)~,
\label{a1} 
\end{equation}
where $P_{\gamma \to \gamma} (E_0, z)$ is the photon survival probability from the source to us, which is usually written in 
terms of the optical depth $\tau_{\gamma} (E_0, z)$ as
\begin{equation}
P_{\gamma \to \gamma} (E_0, z) = e^{ - \tau_{\gamma} (E_0, z)}~.
\label{a2} 
\end{equation}

The main issue we are going to address is a possible {\it correlation} between the 
$\{\Gamma_{\rm em} \}$ {\it distribution} of the blazars in ${\cal S}$ and their {\it redshift} $z$.

At first sight, the reader might well wonder about such a question. After all, why should a correlation of this kind exist for the blazars belonging to ${\cal S}$? Cosmological evolutionary effects in the sources are certainly harmless up to redshift $z = 0.6$ and when observational selection biases are properly taken into account we are unable to figure out which kind of physical mechanism can be responsible for such a statistical correlation. 
 As a consequence, we expect that the best-fit regression line of the $\{\Gamma_{\rm em} \}$ distribution should be a {\it straight horizontal} line in the $\Gamma_{\rm em} - z$ plane.

Actually -- in order to avoid repetitions -- we proceed to summarize the story told in the present paper in a way that also the content of all Sections becomes apparent. 

We report all observational information needed for our treatment in Sect. 2 and Table I of the SM. Sect. 3 is devoted to inferring for any source in ${\cal S}$ the emitted spectral index -- to be denoted by $\Gamma_{\rm em}^{\rm CP} (z)$ -- starting from the observed one $\Gamma_{\rm obs} (z)$ assuming conventional physics (CP), namely taking into account the effect of the EBL absorption alone. After performing a statistical analysis of the emitted spectral indices $\{\Gamma_{\rm em}^{\rm CP} (z) \}$, we end up with the conclusion that the best-fit regression line of the $\{\Gamma_{\rm em}^{\rm CP} (z) \}$ distribution is a concave parabola in the $\Gamma_{\rm em}^{\rm CP} - z$ plane {\it decreasing} as $z$  increases, thereby implying that blazars with harder spectra are found on average {\it only} at larger redshifts. More generally, a {\it statistical correlation} between the $\{\Gamma_{\rm em}^{\rm CP} (z) \}$ distribution and $z$ shows up. Manifestly, the most natural explanation that comes to mind are selection effects. Yet, a deeper scrutiny -- reported in Sect. 4 -- based on the observational information shows that, although the dimming bias can never be avoided, we strongly argue that it does not give rise to the above correlation. Moreover, we demonstrate that the volume dimming is not responsible for such a correlation either. 

So, as explained above, we fail to understand the physical origin of such a conclusion.  
Otherwise stated, how can the sources get to know their $z$ so as to tune their $\Gamma_{\rm em}^{\rm CP} (z)$ in such a way to reproduce the above statistical correlation? We call the existence of such a correlation the {\it VHE BL Lac spectral anomaly}, which of course concerns flaring BL Lac alone. 

In Sect. 5 we turn the argument around, plotting the same data for all sources of ${\cal S}$ in the $\Gamma_{\rm em}^{\rm CP} - z$ plane and {\it imposing by hand} that they are fitted by a horizontal straight line. But we must conclude that such a procedure does not work. 

As an attempt to get rid of the VHE BL Lac spectral anomaly, in Sect. 6 we introduce axion-like particles (ALPs) (for a review, see~\citealt{alp1,alp2}). They are spin zero, neutral and extremely light pseudo-scalar bosons predicted by several extensions of the Standard Model of particle physics, and especially by those based on the M theory which encompasses superstrings and superbranes (see e.g.~\citealt{cisterna2} and references therein). They interact in particular with two photons. Depending on their mass and two-photon coupling, they can be quite good candidates for cold dark matter~\citep{arias} and are attracting an ever growing interest (see~\citealt{gr2018a,gr2018b,gtl2019,gRew} and references therein).

As a matter of fact, we suppose that photon-ALP oscillations~\citep{sikivie,anselmo,rs1988} take place in the extragalactic magnetic field, as first proposed in~\citet{drm2007}. As a consequence, photon propagation gets affected by EBL plus photon-ALP oscillations. Assuming a specific set of allowed values for the model parameters, we re-derive for every source in ${\cal S}$ the emitted spectral index -- to be denoted by $\Gamma_{\rm em}^{\rm ALP} (z)$ -- starting from the observed one $\Gamma_{\rm obs} (z)$. Proceeding then again as before, a statistical analysis of the emitted spectral indices $\{\Gamma_{\rm em}^{\rm ALP} (z) \}$ shows that now the best-fit regression line of the $\{\Gamma_{\rm em}^{\rm ALP} (z) \}$ distribution becomes {\it independent} of redshift, being a {\it horizontal straight line} in the $\Gamma_{\rm em}^{\rm ALP} - z$ plane. This result looks astonishing to us, since -- among infinitely many possibilities -- it is the {\it only one} for which no statistical correlation of the considered kind shows up, and so in agreement with our expectation. We stress that -- if this result were instead due to dimming biases -- then the combined energy sensitivity threshold of H.E.S.S., MAGIC and VERITAS should indeed be  exceptional. Moreover, we show in Sect. 7 that a new scenario for flaring BL Lacs  naturally emerges with the introduction of ALPs, wherein the values of $\Gamma_{\rm em}^{\rm ALP} (z)$ for the individual blazars exhibit a small scatter about the horizontal straight best-fit regression line. Finally, in Sect. 8 we offer our conclusions. It looks amazing  that for the same values of the model parameters which lead to the above result also an explanation arises as to why flat spectrum radio quasars (FSRQs) emit up to 400 GeV, in sharp disagreement with conventional physics (which would predict a gamma-ray emission only up to $30 \, {\rm GeV}$ at most) provided that the emitting region is before or inside the broad line region~\citep{trgb2012}. Thus, the combination of the two very different results -- taken at face value -- offers a hint at an ALP with mass $m = {\cal O} (10^{-10} \, {\rm eV})$ and two-photon coupling in the range $2.94 \times 10^{- 12} \, {\rm GeV}^{- 1} < g_{a \gamma \gamma} < 0.66 \times 10^{ - 10} \, {\rm GeV}^{- 1}$. And as a bonus, the Universe becomes considerably more transparent for $E_0 \gtrsim 1 \, {\rm TeV}$~\citep{dgr2011} as compared to the predictions of conventional physics~\citep{dgr2013}. We conclude Sect. 8 by emphasizing that the ALP in question is planned to be searched for both with the next generation of VHE observatories and in laboratory experiments with various techniques.

In order to avoid breaking the main line of development by long Tables concerning the properties of all considered blazars and somewhat involved technicalities, we report all of them in the Supplementary Material (SM). Specifically, the properties of the blazars considered are reported in Appendix 1 of the SM, the evaluation of the photon survival probability in the presence of ALPs is contained in Appendix 2 of the SM, and Tables containing the emitted properties of all considered blazars inferred within the ALP scenario are listed in Appendix 3 of the SM. Likewise, the counterpart of Sect. 4 for the ALP scenario is presented in Appendix 4 of the SM, whereas an aspect concerning the $z$-dependence of the spectral indices in both contexts is the subject of Appendix 5 of the SM.

We would like to emphasize that the analysis presented in this paper should be regarded as preliminary, and so a more thorough treatment is needed to assess our conclusions. 

\section{OBSERVATIONAL INFORMATION}

\begin{figure}
\centering
\includegraphics[width=.5\textwidth]{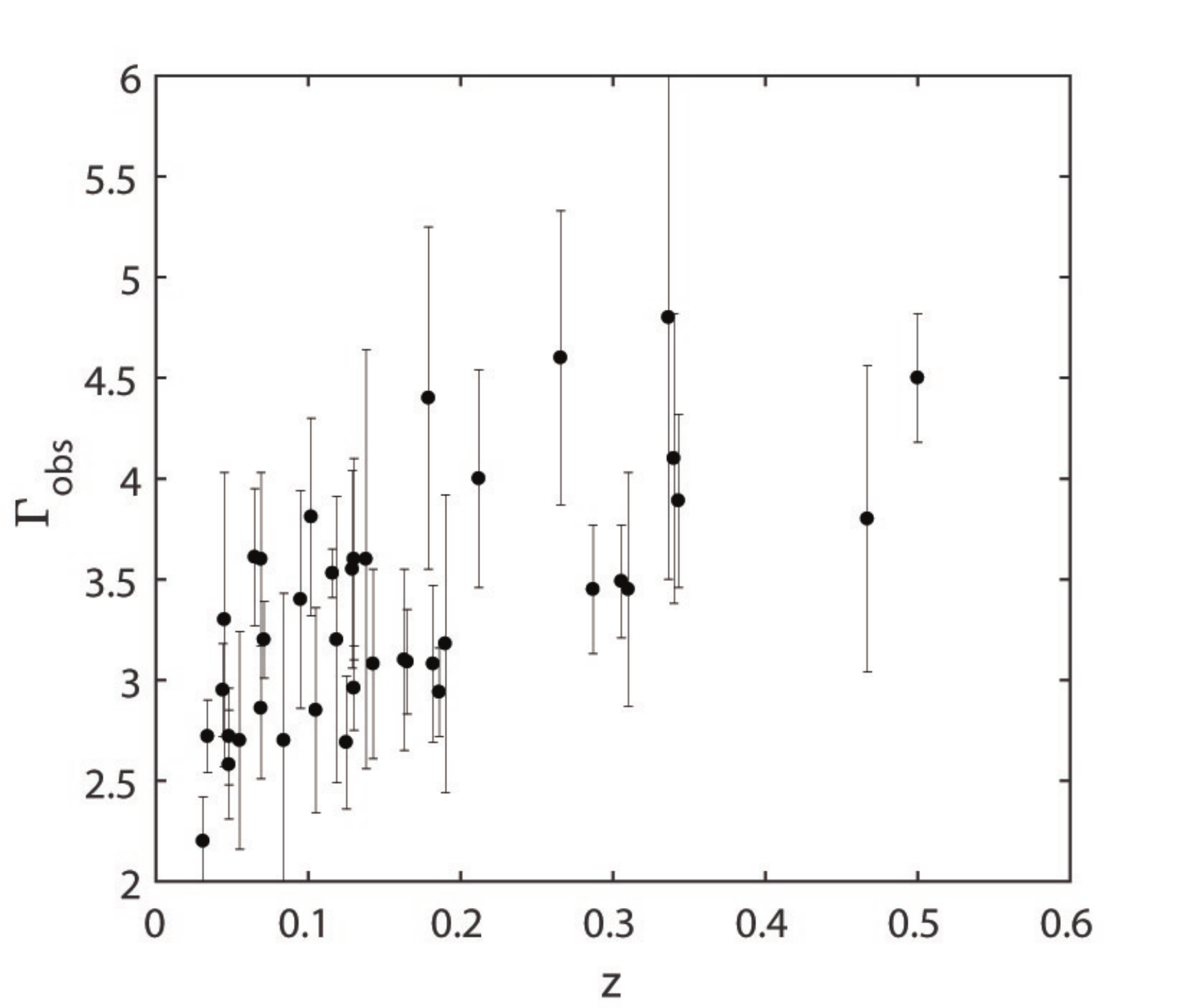}
\caption{\label{fig1} The values of the observed spectral index $\Gamma_{\rm obs}$ with the corresponding error bars are plotted versus $z$ for all blazars in ${\cal S}$.}
\end{figure}

As already pointed out, the observational quantities concerning every blazar in ${\cal S}$ are: the redshift $z$, the observed spectral index $\Gamma_{\rm obs} (z)$ with the associated error bar and the energy range $\Delta E_0 (z)$ wherein a source at redshift $z$ is observed. Specifically, the redshift is taken directly from the Tevcat catalog\footnote{http://Tevcat.uchicago.edu/}, while the other information from the original papers quoted by Tevcat for every blazar (when more than one redshift is available, we use the one reported by Tevcat). All these quantities are summarized in Table I of the SM, and the values of the observed spectral indices $\Gamma_{\rm obs} (z)$ are plotted in Fig.~\ref{fig1} for all sources in ${\cal S}$. 

Moreover, in order to clarify a situation that we shall encounter, we will also need two further quantities. 

One is the observed {\it flux normalization constant} $\hat{K}_{\rm obs} (z)$ entering the expression of $\Phi_{\rm obs}(E_0, z)$ for every source in ${\cal S}$ (we take it from the original papers). Unfortunately, the error bars associated with $\hat{K}_{\rm obs} (z)$ are not always reported in the published papers, but for our considerations this is not a problem. It has to be noted that the quantity $\hat{K}_{\rm obs} (z)$ is generally defined at different energy for different sources. So, for the sake of comparison among all observed flux normalization constants we need to perform a rescaling $\hat{K}_{\rm obs} (z) \to K_{\rm obs} (z)$ in the observed flux of the considered blazars in such a way that $K_{\rm obs} (z)$ coincides with $\Phi_{\rm obs}(E_0, z)$ at the fiducial energy $E_{0,*} = 300 \, {\rm GeV}$ for every source in ${\cal S}$ (this convention is implicitly assumed henceforth throughout the paper). Correspondingly, the observed flux takes the form
\begin{equation}
\label{a417072017}
\Phi_{\rm obs} \left(E_0, z \right) = K_{\rm obs} (z) \left(\frac{E_0}{E_{0,*}} \right)^{- \Gamma_{\rm obs}(z)}~.
\end{equation}

The other quantity is the {\it observed surface luminosity density} $F_{{\rm obs}, \Delta E_0} (z)$, defined as the observed luminosity per unit detector area, which -- recalling that $\Phi_{\rm obs}(E_0, z) = d N_{\rm obs}/(d A \, d t \, d E_0)$ -- clearly arises as the integral of $\Phi_{\rm obs}(E_0, z)$ over the energy range $\Delta E_0 $ for any source belonging to ${\cal S}$, to wit 
\begin{equation}
\label{25022019b}
F_{{\rm obs}, \Delta E} (z) \equiv K_{\rm obs} (z) \int_{E_{0,{\rm min}} (z)}^{E_{0,{\rm max}} (z)} d E_0 ~ \left(\frac{E_0}{E_{0,*}} \right)^{- \Gamma_{\rm obs} (z)}~,
\end{equation}
where $E_{0,{\rm min}} (z)$ and $E_{0,{\rm max}} (z)$ are the lower and upper values 
of $\Delta E_0 (z)$.
The numerical values of $F_{{\rm obs}, \Delta E} (z)$ are obtained from Eq. (\ref{25022019b}) in terms of those reported in Table I of the SM: they are plotted in Fig. \ref{cutoff} below. 

\section{CONVENTIONAL PROPAGATION IN EXTRAGALACTIC SPACE}

After a long period of uncertainty on the EBL precise spectral energy distribution and photon number density, today a convergence has been reached~\citep{ebl}, well represented e.g. by the model of Franceschini and Rodighiero (FR),~\citealt{fr}), which we use for convenience. 

Owing to the $\gamma + \gamma \to e^+ + e^-$ scattering off EBL photons~\citep{bw,heitler}, the emitted VHE photons undergo an energy-dependent absorption, so that the VHE photon survival probability is given by Eq. (\ref{a2}) with $\tau_{\gamma} (E_0, z) \to \tau_{\gamma}^{\rm FR} (E_0, z)$, where $\tau_{\gamma}^{\rm FR} (E_0, z)$ is the optical depth of the EBL as evaluated within the FR model in a standard fashion using the photon spectral number density~\citep{nikishov,gould,fazio}. As a consequence, the observed flux $\Phi_{\rm obs} (E_0, z)$ is related to the emitted one $\Phi^{\rm CP}_{\rm em} (E)$ by 
\begin{equation}
\label{a3}
\Phi_{\rm obs}(E_0,z) = e^{- \tau_{\gamma}^{\rm FR} (E_0, z)} \, \Phi^{\rm CP}_{\rm em} \bigl(E_0 (1+z) \bigr)~,
\end{equation}
where we recall that CP stands for conventional physics. 

\begin{figure}
\begin{center}
\includegraphics[width=.50\textwidth]{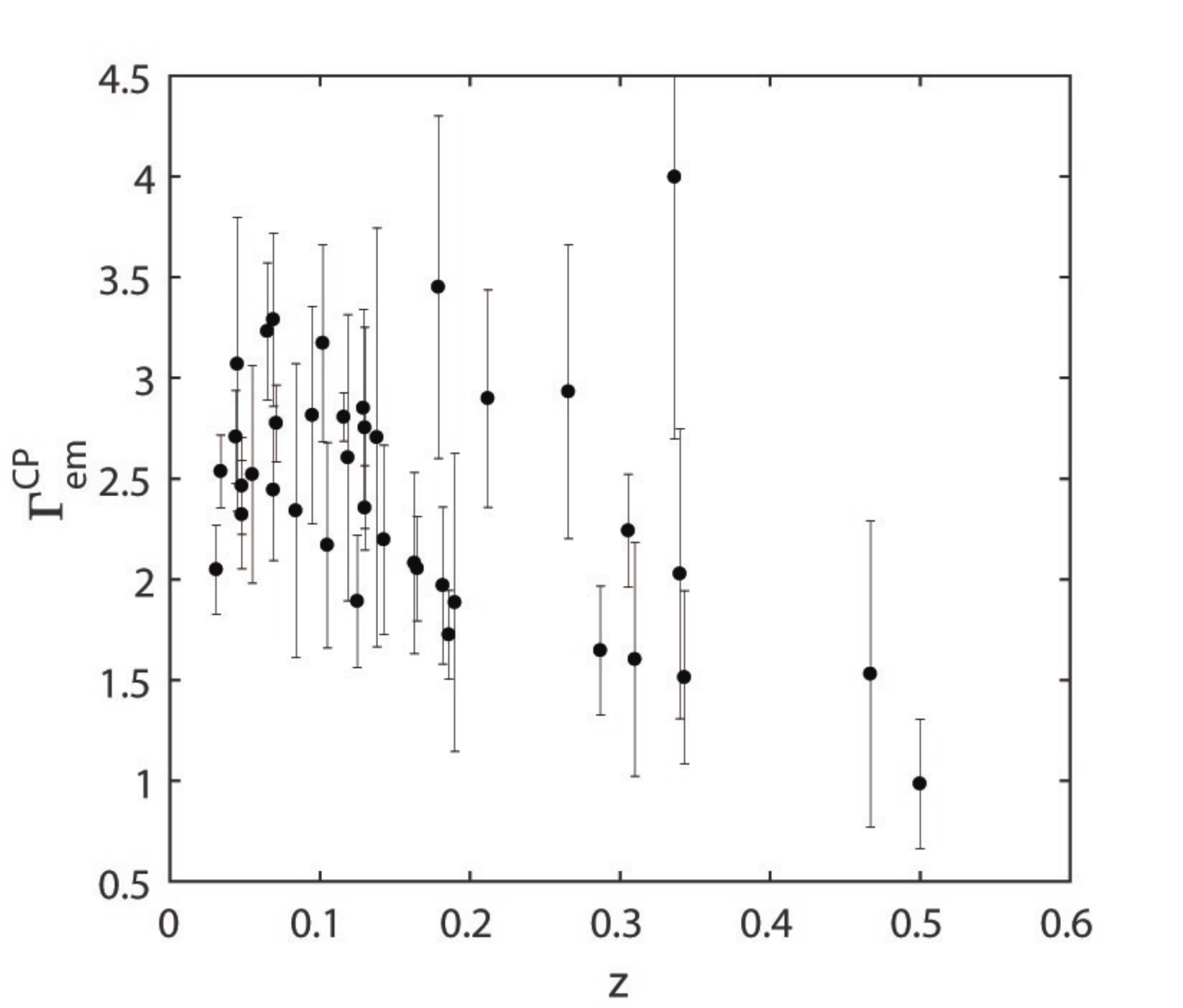}
\end{center}
\caption{\label{fig2a} The values of the emitted spectral index $\Gamma_{\rm em}^{\rm CP}$ with the corresponding error bars are plotted versus $z$ for all blazars in ${\cal S}$.}
\end{figure}

Let us begin by deriving the emitted spectrum of every source in ${\cal S}$, starting from each observed one, within conventional physics. As a preliminary step, thanks to Eq. (\ref{a417072017}) we rewrite Eq. (\ref{a3}) as 
\begin{equation}
\label{a4}
\Phi^{\rm CP}_{\rm em} \bigl(E_0 (1+z) \bigr) = e^{\tau_{\gamma}^{\rm FR} (E_0, z)} \, K_{\rm obs} (z) \left(\frac{E_0}{E_{0,*}} \right)^{- \Gamma_{\rm obs}(z)}~.
\end{equation}
Because of the presence of the exponential in the r.h.s. of Eq. (\ref{a4}), $\Phi^{\rm CP}_{\rm em} \bigl(E_0 (1+z) \bigr)$ cannot behave as an exact power law. Yet, we have pointed out in Sect. 1 that it is expected to be close to it. Therefore, we best-fit (BF) $\Phi^{\rm CP}_{\rm em} \bigl(E_0 (1+z) \bigr)$ to a single power-law expression   
\begin{equation}
\label{a418}
\Phi_{\rm em}^{\rm CP, BF} \bigl(E_0 (1+z) \bigr) = K_{\rm em}^{\rm CP} (z) \, 
\left(\frac{E_0 (1+z)}{E_{0,*}} \right)^{- \Gamma_{\rm em}^{\rm CP} (z)}
\end{equation} 
over the energy range $\Delta E_0 (z)$ where a source is observed, and so $E_0$ varies inside $\Delta E_0 (z)$ (which changes from source to source). Correspondingly, the resulting values of $\Gamma_{\rm em}^{\rm CP} (z)$ and $K_{\rm em}^{\rm CP} (z)$ are quoted in Table II of the SM, and those of $\Gamma_{\rm em}^{\rm CP} (z)$ are plotted in Fig.~\ref{fig2a}. 

\begin{figure}
\begin{center}
\includegraphics[width=.50\textwidth]{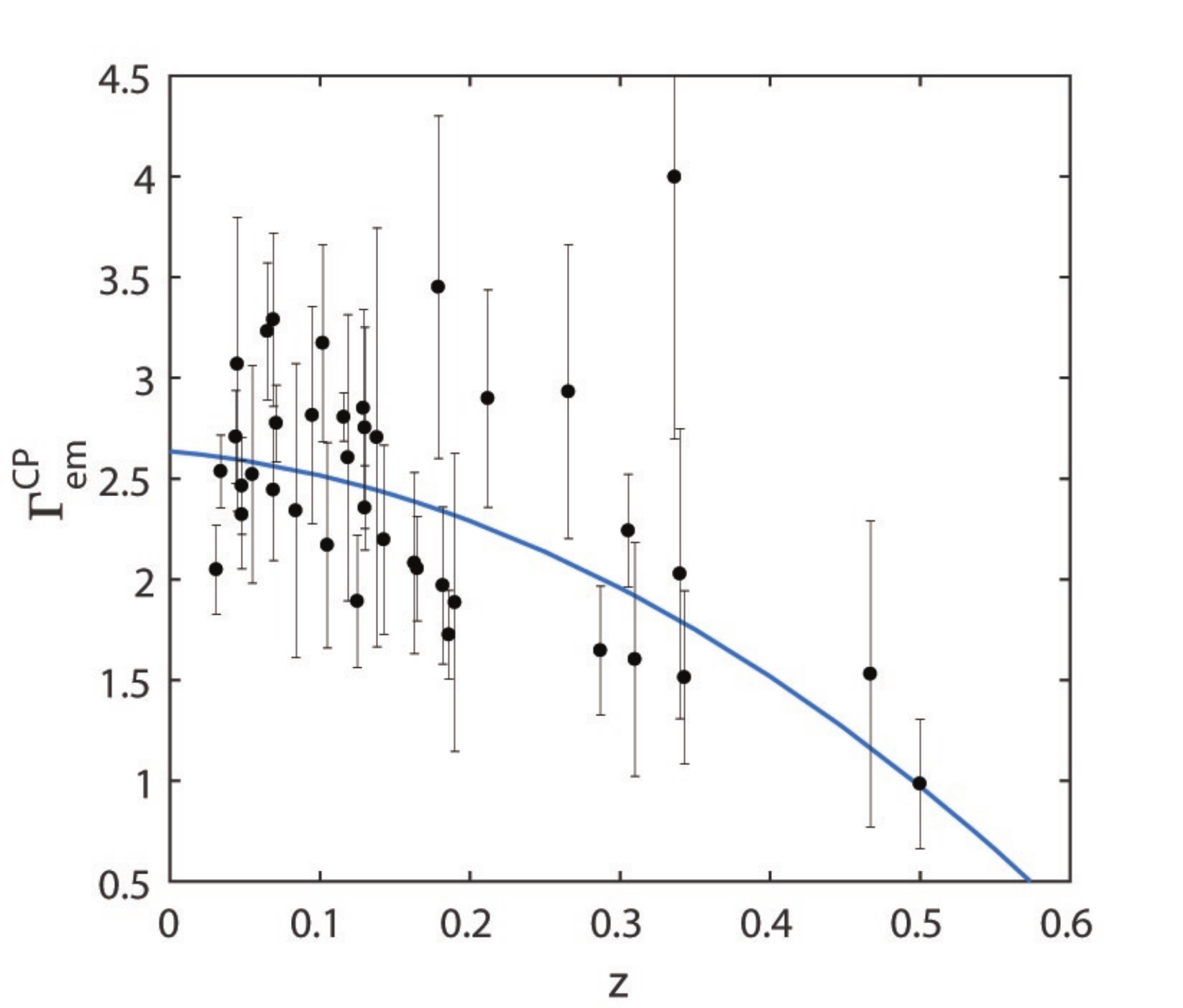}
\end{center}
\caption{\label{fig2b} Same as Fig.~\ref{fig2a}, but with superimposed the best-fit regression line with $\chi^2_{\rm red,CP} = 1.46$.}
\end{figure}

We proceed by performing a statistical analysis of all values of $\Gamma_{\rm em}^{\rm CP} (z)$ as a function of $z$. Specifically, we use the least square method and try to fit the data with one parameter (horizontal straight line), two parameters (first-order polynomial), and three parameters (second-order polynomial). In order to test the statistical significance of the fits we evaluate the corresponding $\chi^2_{\rm red, CP}$. The values of the $\chi^2_{\rm red, CP}$ obtained for the three fits are $2.37$ (one parameter), $1.49$ (two parameters) and $1.46$ (three parameters). Thus, data appear to be best-fitted by the second-order polynomial 
\begin{equation}
\label{a631}
\Gamma_{\rm em}^{\rm CP} (z) = - \, 5.33 \, z^2  - \, 0.66 \, z + 2.64~. 
\end{equation} 
The best-fit regression line given by Eq. (\ref{a631}) turns out to be a concave parabola shown in Fig.~\ref{fig2b}.   

We emphasize that in order to appreciate the physical consequences of Eq. (\ref{a631}) we should keep in mind that $\Gamma_{\rm em}^{\rm CP} (z)$ is the {\it exponent} of the emitted energy entering  $\Phi_{\rm em}^{\rm CP} (E)$. Hence, in the two extreme cases $z = 0$ and $z = 0.6$ we have
\begin{equation}
\label{a412}
\Phi_{\rm em}^{\rm CP} (E, 0) \propto E^{ - 2.64}~, \ \ \ \ \ \ \ \ \ \ \ \ \Phi_{\rm em}^{\rm CP} (E, 0.6) \propto E^{ - 0.33}~,
\end{equation}
thereby implying that the hardening of the emitted flux progressively {\it increases} with the redshift.

The mismatch between our expectation -- namely a straight horizontal line in the $\Gamma_{\rm em} 
- z$ plane -- and the concave parabolic behaviour of the best-fit regression line (\ref{a631}) reported in Fig.~\ref{fig2b} is termed {\it VHE BL Lac spectral anomaly}.

\section{SELECTION BIASES}

As previously stressed -- since we are concerned with a relatively local sample -- cosmological evolutionary effects in the sources are insignificant (observe that evolutionary effects in the considered EBL model have instead been taken into account).

Therefore, {\it a priori} the most obvious origin of the VHE BL Lac spectral anomaly is expected to be a selection bias. Hence, a very careful analysis of this issue is compelling. 

A quite instrumental quantity in this respect -- defined in analogy with Eq. (\ref{25022019b}) -- is the {\it emitted surface luminosity density} $F_{{\rm em}, \Delta E}^{\rm CP} (z)$, defined as the emitted luminosity per unit emitting area, which -- recalling that $\Phi_{\rm em} (E) = d N_{\rm em}/(d A \, d t \, d E)$ -- manifestly arises as the integral of $\Phi_{\rm em} (E)$ over $\Delta E (z) = (1 + z) \Delta E_0 (z)$ for any source in ${\cal S}$, namely
\begin{eqnarray}
\label{17012019a}
&\displaystyle F_{{\rm em}, \Delta E}^{\rm CP} (z) \equiv (1 + z) \, K_{\rm em}^{\rm CP} (z) \times \\
&\displaystyle \times \int_{E_{0,{\rm min}} (z)}^{E_{0,{\rm max}} (z)} d E_0 ~
\left(\frac{E_0 (1 + z)}{E_{0,*}} \right)^{- \Gamma_{\rm em}^{\rm CP} (z)}~, \nonumber
\end{eqnarray}
where $E_{0,{\rm min}} (z)$ and $E_{0,{\rm max}} (z)$ are the lower and upper values of the energy range $\Delta E_0 (z)$. Both the values of $\Delta E_0 (z)$ and those of $F_{{\rm em}, \Delta E}^{\rm CP} (z)$ are reported in Table II of the SM.

\

Basically, two selection biases are relevant in the present context.

\ 

\noindent {\bf Dimming bias:} {\it As we look at larger distances only the brighter sources are observed while the fainter ones progressively disappear}. This arises from two distinct effects. 

\medskip 

\noindent {\it Malmquist bias} -- This is due to the fact that the really emitted luminosity goes like the inverse of the square distance. As already stressed, it does not affect the shape of blazar spectra and so it cannot be responsible for the VHE BL Lac spectral anomaly. Hence, it will henceforth be discarded.

\medskip 

\noindent {\it EBL-absorption} -- This is due to the reaction $\gamma + \gamma \to e^+ + e^-$, which strongly affects the spectra.

\

\noindent {\bf Volume bias:} {\it Looking at greater distances entails that larger regions of space are probed, and so -- under the assumption of an uniform source distribution -- a larger number of brighter blazars should be detected}. Because this is a purely geometric argument, the EBL-absorption must be subtracted. As a consequence, we have to deal with deabsorbed spectra, namely with the {\it emitted} ones. 

\bigskip 

\subsection{Discussion of the dimming bias}

\medskip

\begin{figure}
\begin{center}
\includegraphics[width=.50\textwidth]{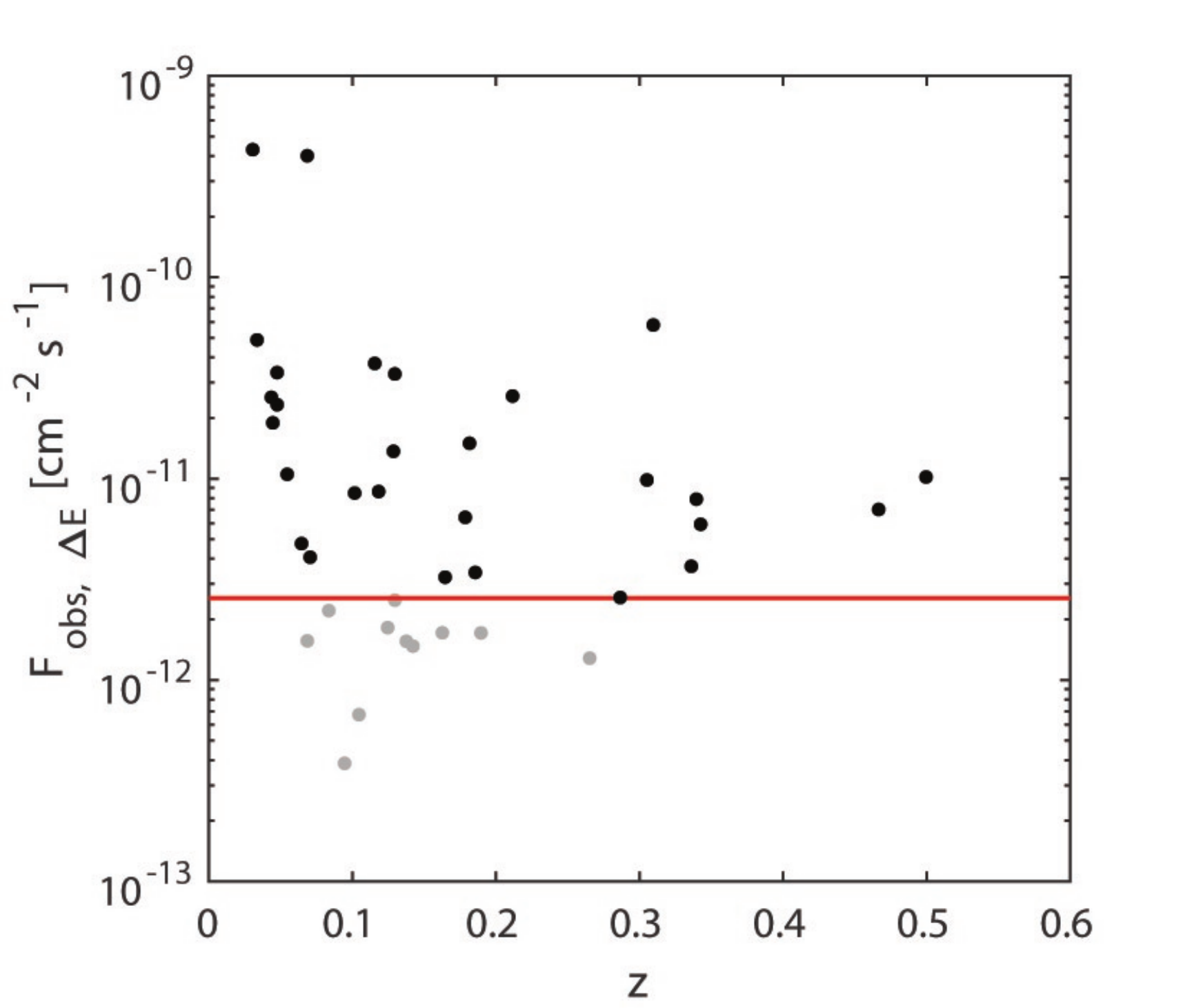}
\end{center}
\caption{\label{cutoff} Out of the original sample ${\cal S}$ -- represented by all points -- we have constructed the subsample ${\cal S}_0$ of sources denoted by black points.}
\end{figure}

It is obvious from the outset that nobody can get rid of the EBL-absorption in general, and the only way to reduce it is to lower the instrumental energy sensitivity threshold. Yet, both in order to probe its relevance -- as well as to check how stable is the result of Sect. 3 against a change in the number of sources -- we proceed as follows. We somewhat arbitrarily divide the considered $z$ range into two equal parts, the midpoint being $z = 0.3$. Then, we remove the 11 sources of ${\cal S}$ with $z < 0.3$. Accordingly, the original sample ${\cal S}$ of 39 sources gets reduced to the subsample ${\cal S}_0$ of 28 sources: this construction is illustrated in Fig~\ref{cutoff}, where also the corresponding values of $F_{{\rm obs}, \Delta E} (z)$ obtained from Eq. (\ref{25022019b}) -- in terms of the quantities reported in Table I of the SM -- are plotted. 

\begin{figure}
\begin{center}
\includegraphics[width=.50\textwidth]{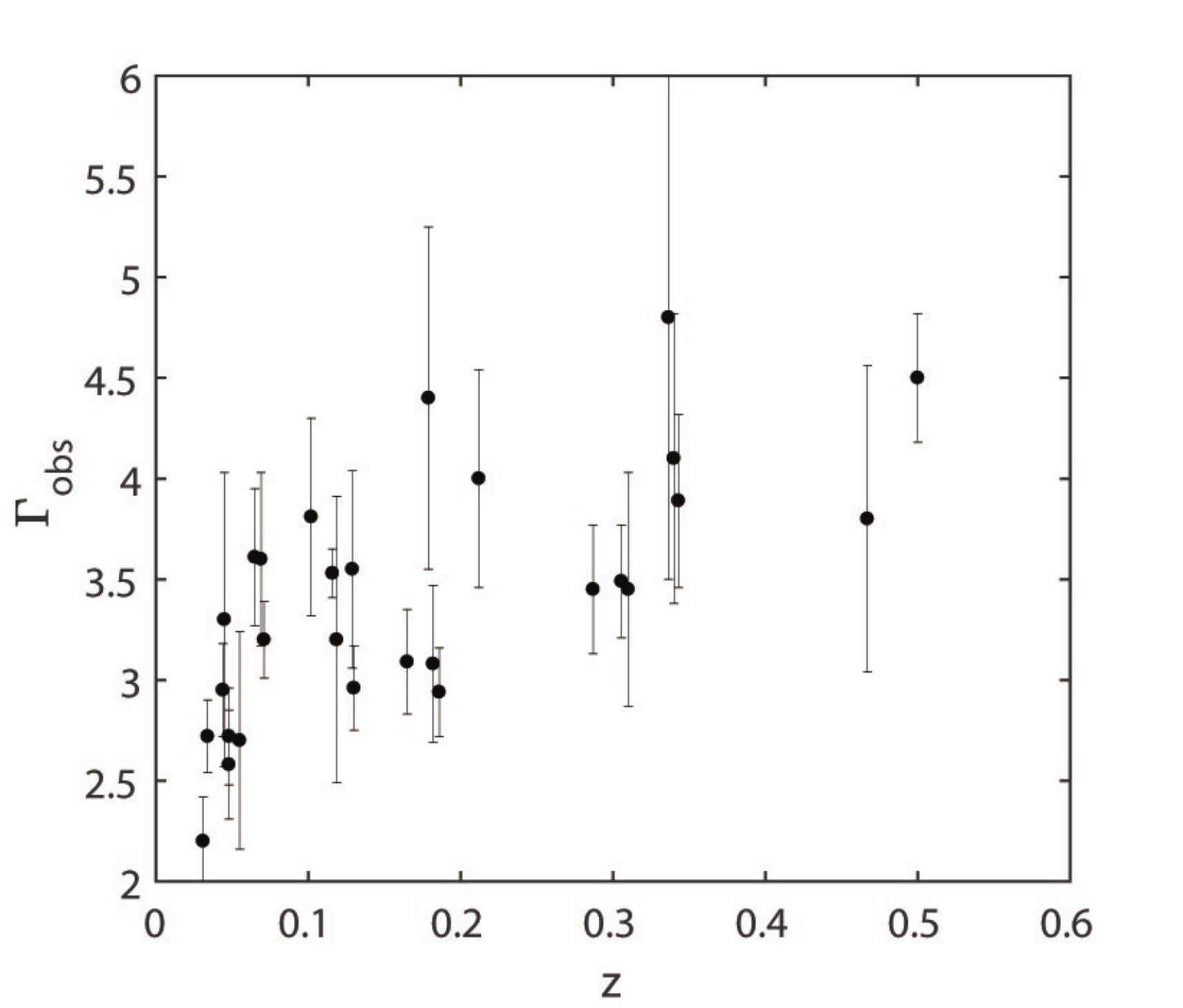}
\end{center}
\caption{\label{slopenoBias} The values of $\Gamma_{\rm obs}$ with the corresponding error bars are plotted versus $z$ for all blazars of ${\cal S}_0$.}
\end{figure}

Moreover, the plot of the values of $\Gamma_{\rm obs} (z)$ pertaining to ${\cal S}_0$ -- listed in Table I of the SM -- is shown in Fig.~\ref{slopenoBias}. 

\begin{figure}
\begin{center}
\includegraphics[width=.50\textwidth]{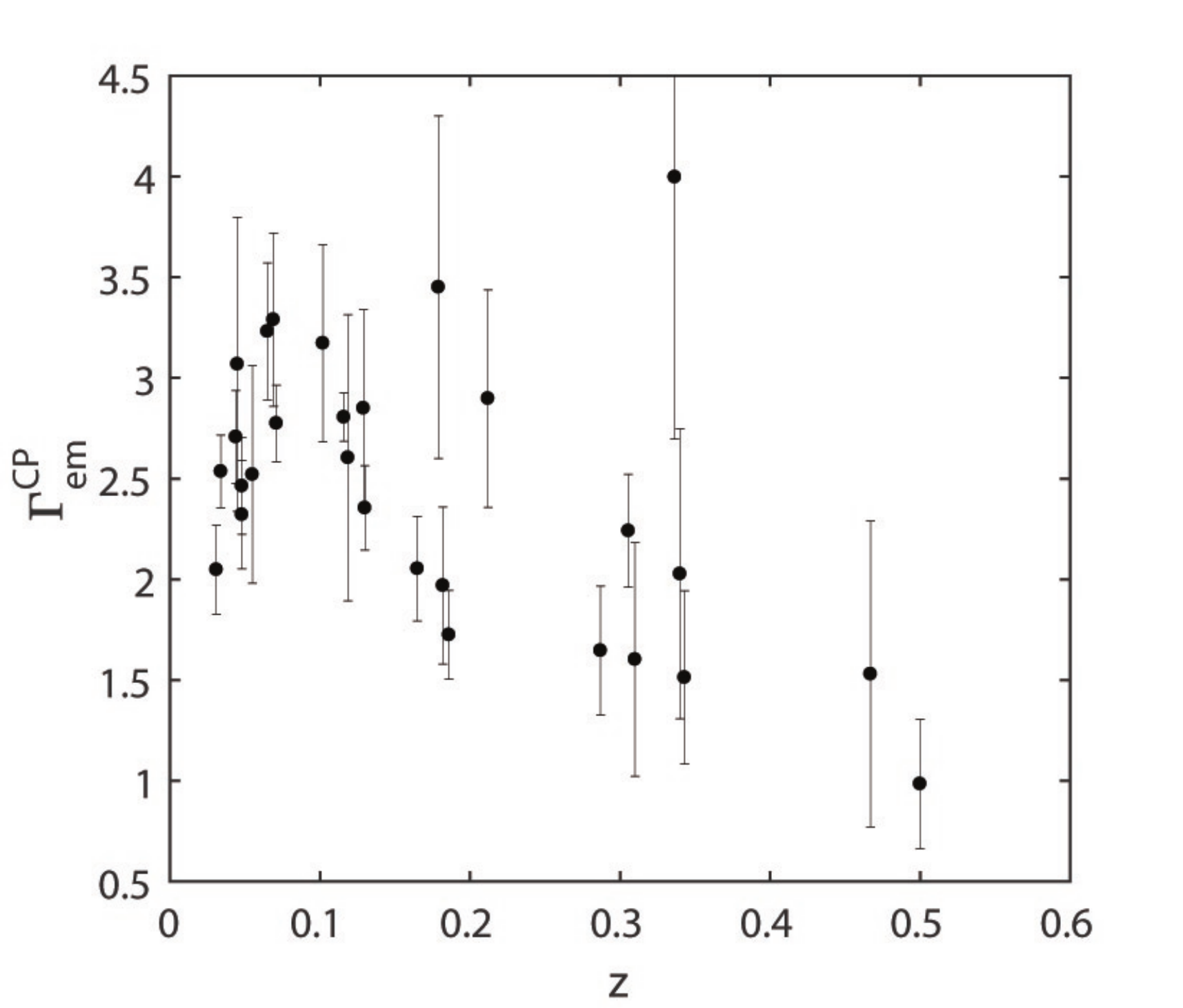}
\end{center}
\caption{\label{fig2noBiasa} The values of $\Gamma_{\rm em}^{\rm CP}$ with the corresponding error bars are plotted versus $z$ for all blazars in ${\cal S}_0$.}
\end{figure}

We proceed along just the same steps of the analysis reported in in Sect. 3. So, we start by de-absorbing the spectra of all sources in ${\cal S}_0$. The resulting values of the emitted spectral index -- reported in Table II of the SM -- are  plotted in Fig.~\ref{fig2noBiasa}.

\begin{figure}
\begin{center}
\includegraphics[width=.50\textwidth]{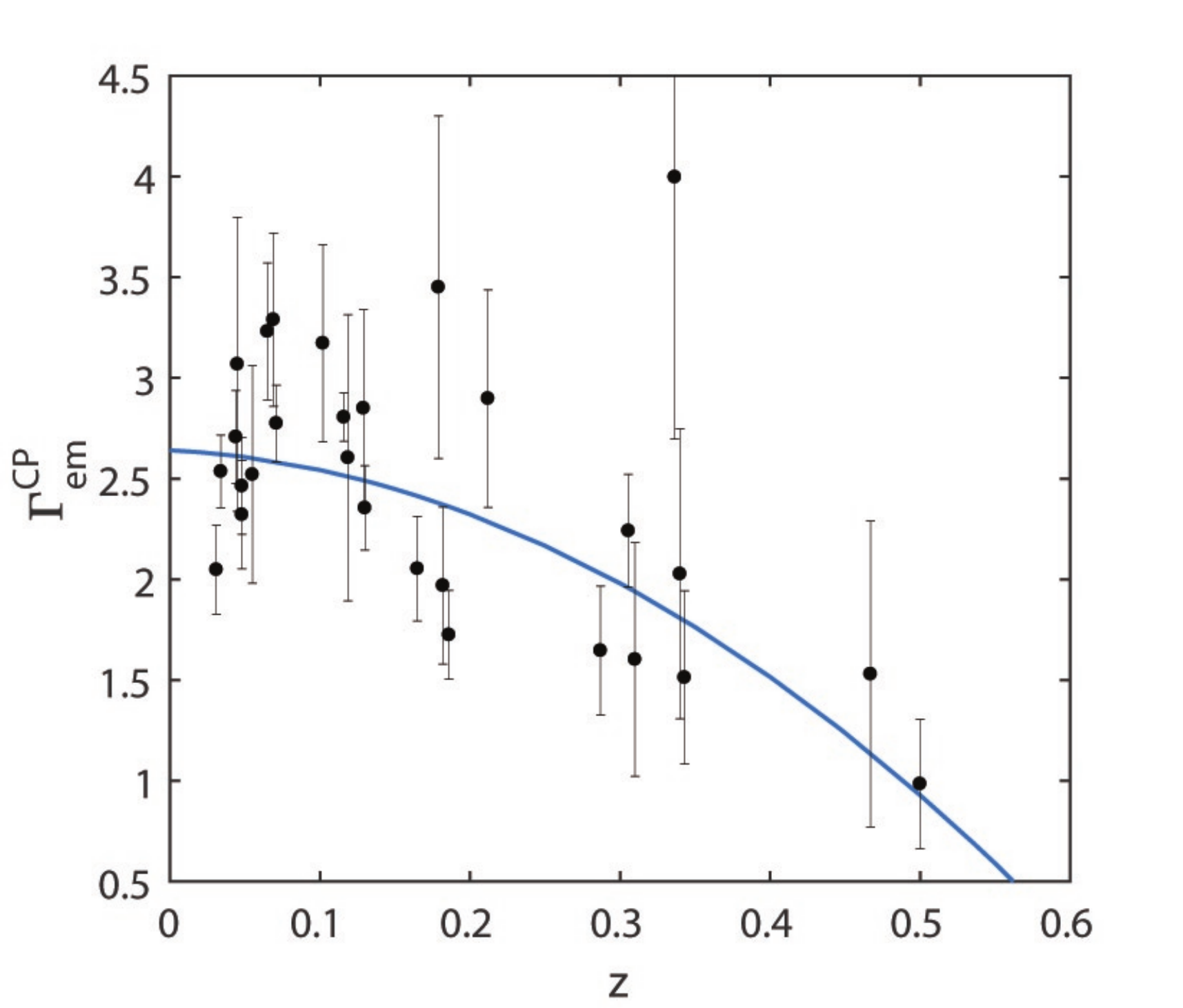}
\end{center}
\caption{\label{fig2noBiasb} Same as Fig.~\ref{fig2noBiasa}, but with superimposed the best-fit regression line with $\chi^2_{\rm red, CP} = 1.82$.}
\end{figure}

Next, we perform a statistical analysis of all values of $\Gamma_{\rm em}^{\rm CP} (z)$ in Fig.~\ref{fig2noBiasa}. As before, we use the least square method and try to fit the data with one parameter (horizontal straight line), two parameters (first-order polynomial), and three parameters (second-order polynomial). In order to test the statistical significance of the fits we evaluate the corresponding $\chi^2_{\rm red, CP}$. The values of the $\chi^2_{\rm red, CP}$ obtained for the three fits are $3.12$ (one parameter), $1.87$ (two parameters) and $1.82$ (three parameters). Thus, data appear to be best-fitted by the second-order polynomial 
\begin{equation}
\label{a631noBias}
\Gamma_{\rm em}^{\rm CP} (z) = - \, 6.11 \, z^2  - \, 0.37 \, z + 2.64~. 
\end{equation} 
The best-fit regression line given by Eq. (\ref{a631noBias}) is shown in  Fig.~\ref{fig2noBiasb}. 

Thus, the result for ${\cal S}_0$ is hardly distinguishable from that for ${\cal S}$: just compare  Fig.~\ref{fig2b} with Fig.~\ref{fig2noBiasb}. So, this exercise {\it strongly suggests} -- even if does not demonstrate -- that the existence of the VHE BL Lac spectral anomaly is quite robust.  

\subsection{Discussion of the volume bias}

\begin{figure}
\begin{center}
\includegraphics[width=.50\textwidth]{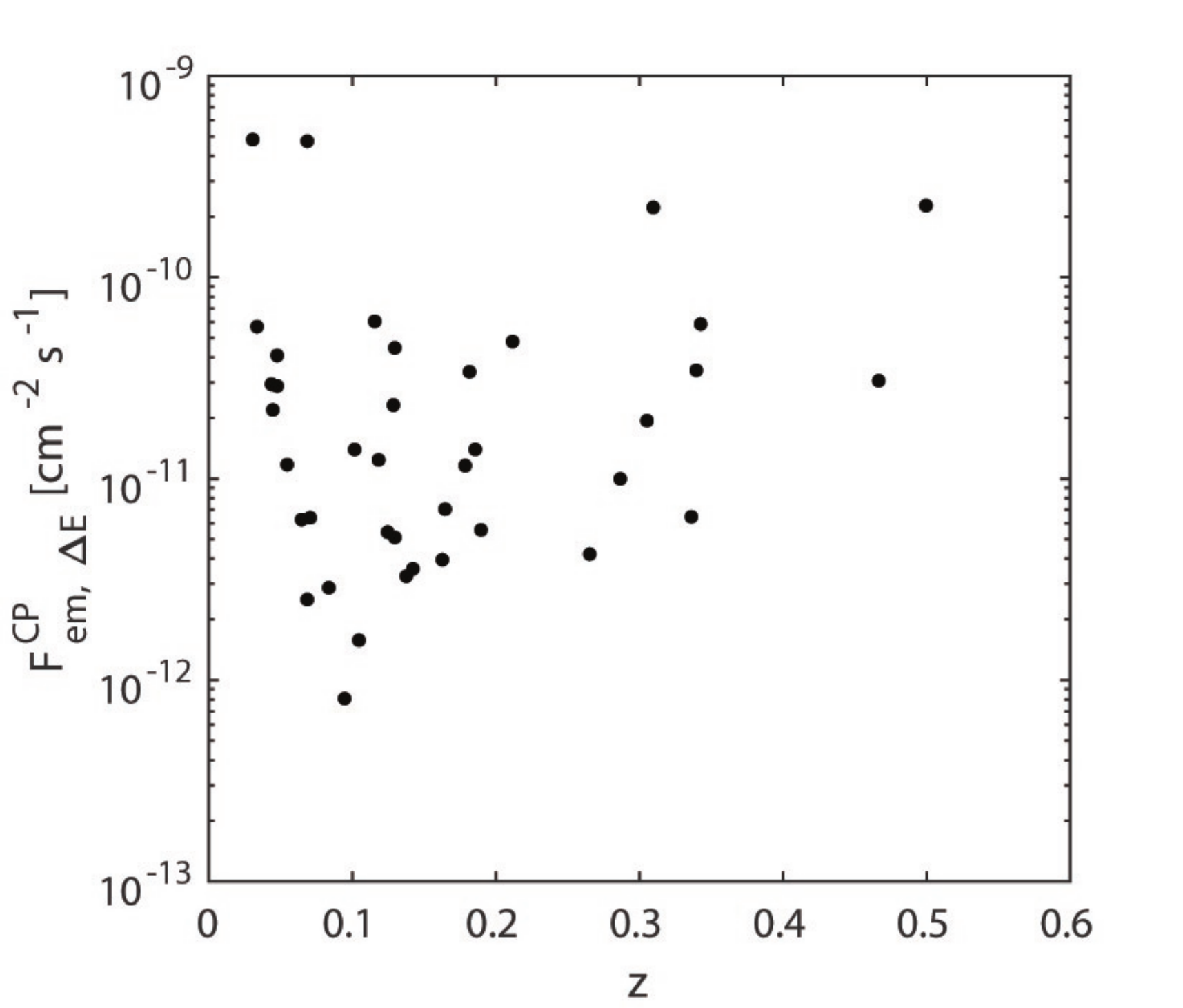}
\end{center}
\caption{\label{figemcp} The values of $F_{{\rm em}, \Delta E}^{\rm CP}$ are plotted versus $z$ for all blazars of ${\cal S}$.}
\end{figure}

Let us come back to our whole blazar sample ${\cal S}$, plotting the values of $F_{{\rm em}, \Delta E}^{\rm CP} (z)$ from Table II of the SM in Fig.~\ref{figemcp}.

A look at Fig.~\ref{figemcp} shows that the volume selection bias is indeed present, thereby providing a check that our procedure is correct. 

Now, the physical explanation of the parabolic best-fit regression line (\ref{a631}) shown in Fig.~\ref{fig2b} would naturally arise only under the presumed {\it assumption} that $F_{{\rm em}, \Delta E}^{\rm CP} (z)$ were {\it tightly correlated} with $\Gamma_{\rm em}^{\rm CP} (z)$ in such a way that {\it brighter sources have harder spectra as} $z$ {\it increases}. Otherwise stated, $F_{{\rm em}, \Delta E}^{\rm CP} (z)$ {\it should increase as} $\Gamma_{\rm em}^{\rm CP} (z)$ {\it decreases and} $z$ {\it increases}. Then the considered selection bias would translate into the statement that looking at greater distances implies that a larger number of blazars with harder spectra should be observed, which is precisely what Fig.~\ref{fig2b} tells us. 

Hence, the real question concerns the validity of the presumed assumption. Now, a glance at Fig.~\ref{figemcp} shows that $F_{{\rm em}, \Delta E}^{\rm CP} (z)$ indeed increases with $z$. Therefore, the previous question becomes: does $F_{{\rm em}, \Delta E}^{\rm CP}$ increase as $\Gamma_{\rm em}^{\rm CP}$ decreases? Actually, the answer is already implicitly contained in Eq. (\ref{17012019a}). In order to bring it out explicitly, we perform the integral. The result is

\begin{eqnarray}
&\displaystyle F_{{\rm em}, \Delta E}^{\rm CP} (z) = \frac{K_{\rm em}^{\rm CP} (z) \, E_{0,*}}{\Gamma_{\rm em}^{\rm CP} (z) - 1} \times \label{09042019a} \\
&\displaystyle \times {\Big [}\left(\frac{E_{0,*}}{E_{0,{\rm min}} (z) (1 + z )} \right)^{\Gamma_{\rm em}^{\rm CP} (z) - 1} - \nonumber \\
&\displaystyle - \left(\frac{E_{0,*}}{E_{0,{\rm max}} (z) (1 + z )} \right)^{\Gamma_{\rm em}^{\rm CP} (z) - 1} {\Big]}~. \nonumber 
\end{eqnarray} 

\medskip

\begin{figure*}
\begin{center}
\includegraphics[width=.50\textwidth]{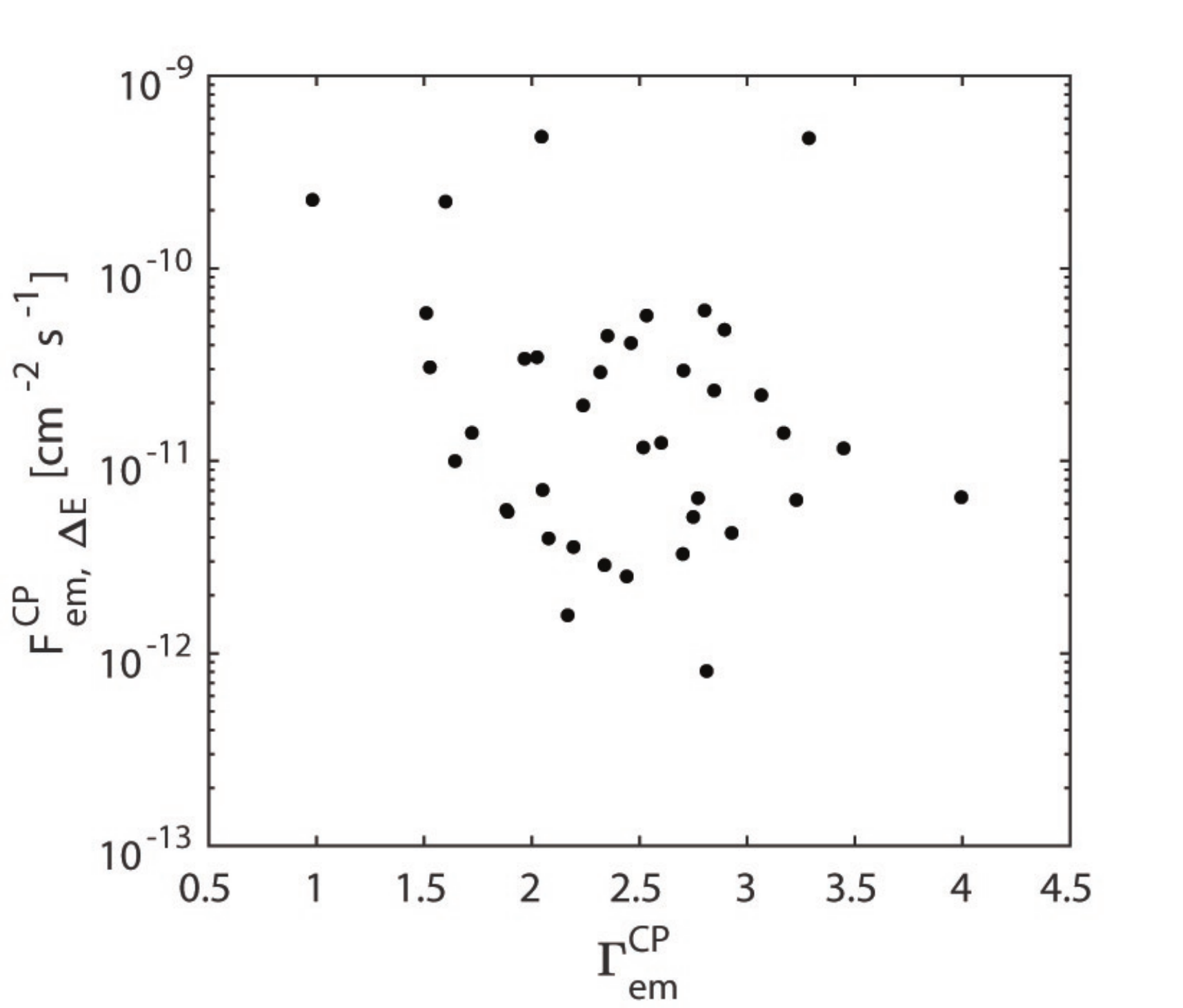}\includegraphics[width=.50\textwidth]{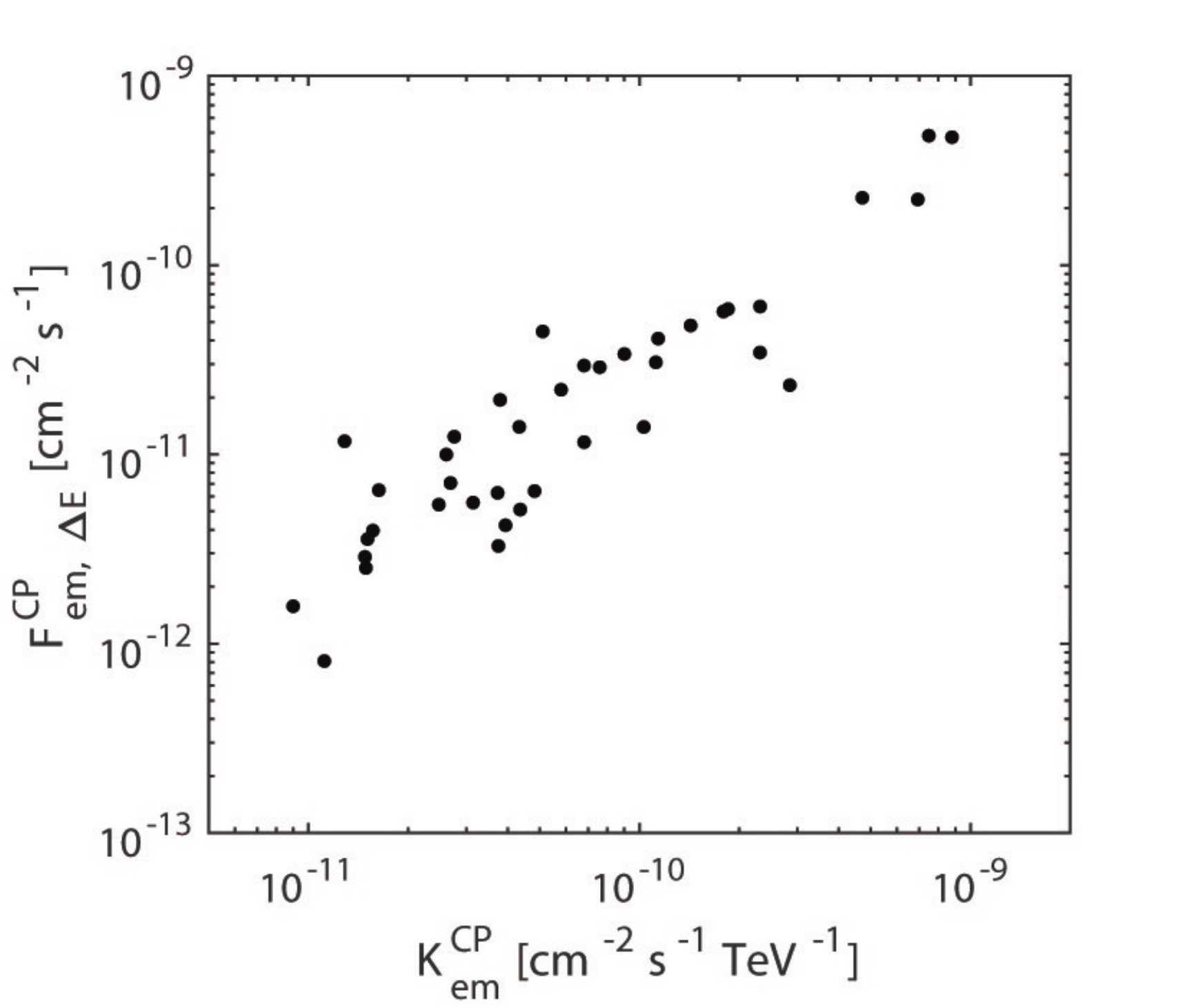} 
\end{center}
\caption{\label{FluxVsSlopecp} {\it Left panel}: the values of $F_{{\rm em}, \Delta E}^{\rm CP}$ are plotted versus $\Gamma_{\rm em}^{\rm CP}$ for all blazars in ${\cal S}$. {\it Right panel}: the values of $F_{{\rm em}, \Delta E}^{\rm CP}$ are plotted versus 
$K_{\rm em}^{\rm CP}$ for all blazars in ${\cal S}$.} 
\end{figure*}

In order to get a deeper insight into the meaning of Eq. (\ref{09042019a}), we note from Table II of the SM that only in the case of PG 1553+113 is $\Gamma_{\rm em}^{\rm CP} < 1$, whereas all other sources have $\Gamma_{\rm em}^{\rm CP} > 1$. So, in order to simplify our discussion we neglect PG 1553+113. Let us consider first the factor in front of the square brackets. Evidently, as $\Gamma_{\rm em}^{\rm CP}$ decreases such a factor increases. What is the effect of the terms in the square brackets? Recalling that $E_{0,*} = 300 \, {\rm GeV}$, a look at Table II of the SM shows that the term involving $E_{0,{\rm max}} (z) \, (1 + z)$ is always larger than $E_{0,*}$, whereas the term containing $E_{0,{\rm min}} (z)$ can be larger or smaller than $E_{0,*}$. In either case, it follows from Table II of the SM that the dominant term in the square bracket is the first. Therefore, when it is larger than 1 the square bracket decreases as $\Gamma_{\rm em}^{\rm CP}$ decreases, while in the opposite case the square bracket increases as $\Gamma_{\rm em}^{\rm CP}$ decreases. By and large, $F_{{\rm em}, \Delta E}^{\rm CP} (z)$ is linearly proportional to $K_{\rm em}^{\rm CP} (z)$, up to random variations around $K_{\rm em}^{\rm CP} (z)$ brought about by $\Gamma_{\rm em}^{\rm CP} (z)$. A picture clarifies the situation: we plot $F_{{\rm em}, \Delta E}^{\rm CP}$ both versus $\Gamma_{\rm em}^{\rm CP}$ as well as versus $K_{\rm em}^{\rm CP}$ in Fig.~\ref{FluxVsSlopecp}.

Therefore, we end up with the conclusion that the assumed correlation -- which would provide a cheap explanation of Fig.~\ref{fig2b} -- is just wrong. 

\bigskip 

\centerline{ \ \ \ * \ \ \ * \ \ \ * \ \ \ }

\medskip

Thus, the behaviour of the parabolic best-fit regression line given by Eq. (\ref{fig2b}) and shown in Fig.~\ref{fig2b} is {\it not due to the volume bias}. Even though we {\it cannot exclude} that the VHE BL Lac spectral anomaly might arise from EBL-absorption our analysis in Subsect. 4.1 has shown that this is a very remote possibility, and anyway cannot arise from the volume bias. Therefore, we are strongly motivated to believe that  conventional physics is afflicted by the VHE BL Lac spectral anomaly.

The redshift dependence of the slope difference $\Gamma_{\rm obs} (z) - \Gamma^{\rm CP}_{\rm em} (z)$ is discussed in Appendix 5 of the SM.

\section{AN ATTEMPT BASED ON CONVENTIONAL PHYSICS}

\begin{figure}     
\centering
\includegraphics[width=0.50\textwidth]{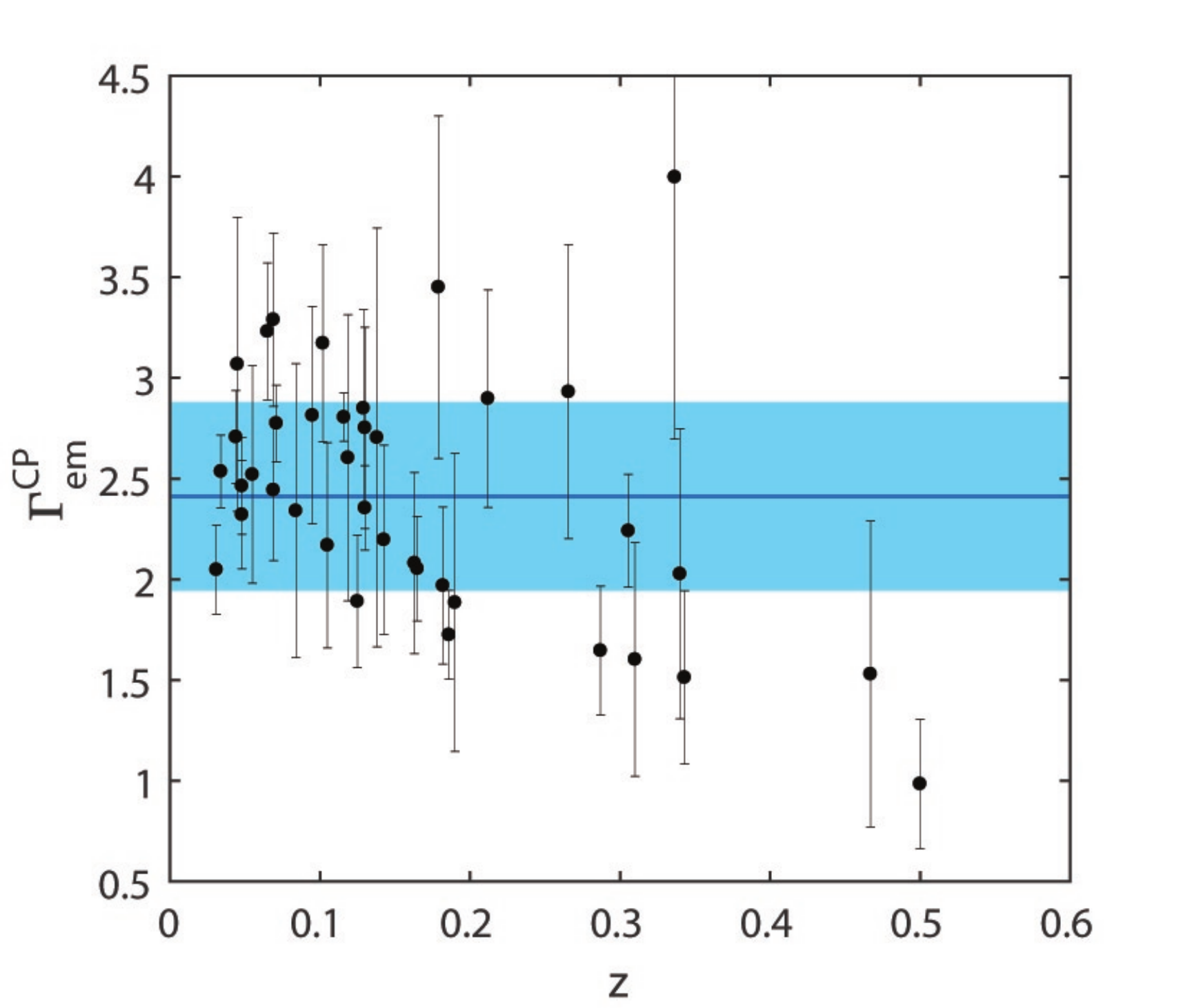}
\caption{\label{fig3} Horizontal fitting straight line in conventional physics. The values of 
$\Gamma_{\rm em}^{\rm CP}$ with the corresponding error bars are plotted versus $z$ for all blazars belonging to $\cal S$. Superimposed is the horizontal straight regression line $\Gamma_{\rm em}^{\rm CP} = 2.41$ with $\chi^2_{\rm red, CP} = 2.37$. The light blue strip encompasses $95 \, \%$ of the sources and its total width is 0.94, which equals $39 \, \%$ of the mean value $\Gamma_{\rm em}^{\rm CP} = 2.41$.}
\end{figure}

In spite of the previous findings, let us nonetheless try to {\it impose by hand} that the same data set $\{\Gamma_{\rm em}^{\rm CP} (z) \}$ corresponding to ${\cal S}$ be fitted by a horizontal straight regression line -- so that the VHE BL Lac spectral anomaly would disappear -- and let us see what happens. Proceeding as before, the result is exhibited in Fig.~\ref{fig3}. In this case, we have $\Gamma_{\rm em}^{\rm CP} = 2.41$ and $\chi^2_{\rm red, CP} = 2.37$. 

Manifestly, this scenario is very hard to believe, since the value of $\chi^2_{\rm red, CP}$ is unduly large. By and large, this fact is to be expected, since we are not best-fitting the data. Still, this is a useful exercise, since it quantifies the price we have to pay in order to have a horizontal straight regression line -- in agreement with our expectation -- within conventional physics, and it is a benchmark for comparison when a new scenario for flaring blazars will be considered in Sect. 7.

Much for the same reason, it is instructive to encompass $95 \, \%$ of the considered sources inside a strip centered on the horizontal fitting line $\Gamma_{\rm em}^{\rm CP} = 2.41$. What is its half-width $\delta \Gamma_{\rm em}^{\rm CP}$? The answer is $\delta \Gamma_{\rm em}^{\rm CP} = 0.47$, which is $20 \, \%$ of the value $\Gamma_{\rm em}^{\rm CP} = 2.41$. This means that the scatter of the values of $\Gamma_{\rm em}^{\rm CP} (z)$ for $95 \, \%$ of the blazars in ${\cal S}$ is at most $20 \, \%$ about the mean value set by horizontal straight regression line, namely 0.47.

\section{AN ATTEMPT BASED ON AXION-LIKE PARTICLES (ALPs)}

\begin{figure}
\begin{center}
\includegraphics[width=.25\textwidth]{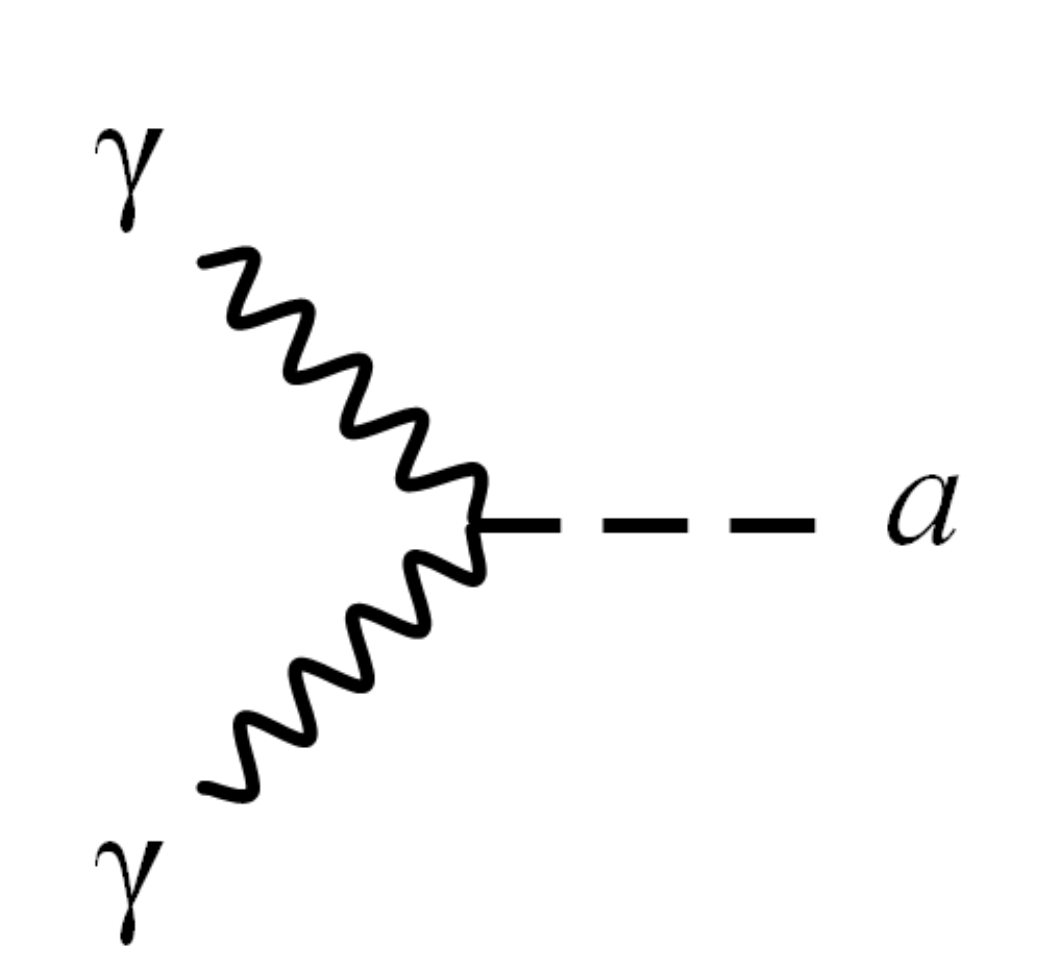}
\end{center}
\caption{\label{immagine3(2)} Feynman diagram for the two-photon ALP coupling.}
\end{figure}

As an alternative possibility to avoid the VHE BL Lac spectral anomaly, we invoke new physics in the form of axion-like particles (ALPs). Their most characteristic feature is to couple to two photons with a coupling constant $g_{a \gamma \gamma}$ according to the Feynman diagram shown in Fig.~\ref{immagine3(2)}. 

Clearly, in the presence of the extragalactic magnetic field ${\bm B}$ one photon line in Fig.~\ref{immagine3(2)} represents the ${\bm B}$ field -- while the other photon line the ${\bm E}$ field of a propagating photon -- and so we see that in such a situation energy-conserving conversions between VHE gamma-rays and ALPs take place. Because these conversions repeatedly occur as a photon/ALP beam emitted by the source propagates towards us, {\it photon-ALP oscillations} show up~\citep{drm2007}. Incidentally, this kind of oscillations are quite similar to those of massive neutrinos of different flavour, apart from the fact that the photon has spin one while the ALP has spin zero, hence the presence of an external field is compelling in order to compensate for the spin mismatch~\citep{sikivie,anselmo,rs1988}. 

Accordingly, photons acquire a split personality, traveling for some time as real photons -- which suffer from EBL-absorption -- and for some time as ALPs, which are unaffected by the EBL (as explicitly shown in Appendix 2 of the SM). Therefore, $\tau_{\gamma} (E_0, z)$ -- chosen before to be equal to $\tau_{\gamma}^{\rm FR} (E_0, z)$ -- gets replaced by the effective optical depth $\tau_{\gamma}^{\rm eff} (E_0, z)$, which is manifestly 
{\it smaller} than $\tau^{\rm FR}_{\gamma} (E_0, z)$ and $\tau_{\gamma}^{\rm eff} (E_0, z)/\tau^{\rm FR}_{\gamma} (E_0, z)$ turns out to be a monotonically decreasing function of both $E_0$ and $z$. Because in the present context Eq. (\ref{a2}) gets replaced by 
\begin{equation}
P_{\gamma \to \gamma}^{\rm ALP} (E_0, z) = e^{ - \tau_{\gamma}^{\rm eff} (E_0, z)}~,
\label{a217072017} 
\end{equation}
the crux of the argument is that even a {\it small} decrease of $\tau_{\gamma}^{\rm eff} (E_0, z)$ with respect to $\tau^{\rm FR}_{\gamma} (E_0, z)$  gives rise to a {\it large} increase of the photon survival probability as compared to the case of conventional physics. So, the main consequence of photon-ALP oscillations is to {\it substantially attenuate} the EBL-absorption, thereby enlarging the cosmic transparency above $E_0 \gtrsim 1 \, {\rm TeV}$. It goes without saying that all this takes place for a specific set of allowed values of the model parameters.

Actually, $P_{\gamma \to \gamma}^{\rm ALP} (E_0, z)$ can be computed exactly as outlined in Appendix 2 of the SM, where it is also explained that the extragalactic magnetic field ${\bm B}$ is usually supposed to originate from quasars and primeval galactic outflows, and it is modeled as a domain-like network, where ${\bm B}$ is homogeneous over a domain of size $L_{\rm dom}$ set by the ${\bm B}$ coherence length and has approximately the same strength $B$ in all domains, but its direction changes randomly from one domain to the next~\citep{kronberg,gr2001}. As we see in Appendix 2 of the SM, the preferred ranges are $1 \, {\rm Mpc} \lesssim L_{\rm dom} \lesssim 10 \, {\rm Mpc}$ and $0.1 \, {\rm nG} \lesssim B \lesssim 1 \, {\rm nG}$, with the upper bound of $1.7 \, {\rm nG}$ at the $2 \sigma$ level~\citep{pshirkov2016}. As far as $g_{a \gamma \gamma}$ is concerned, two observational upper limits are available: one obtained by the CAST experiment at CERN which reads $g_{a \gamma \gamma} < 0.66 \times 10^{- 10} \, {\rm GeV}^{- 1}$ for an ALP mass $m$ obeying the condition $m < 0.02 \, {\rm eV}$ at the $2 \sigma$ level~\citep{cast2017}, and an identical astrophysical one based on the study of 39 Galactic globular clusters~\citep{straniero}. We also see in Appendix 2 that the considered ALP scenario contains three free parameters which enter 
$P_{\gamma \to \gamma}^{\rm ALP} (E_0, z)$: $L_{\rm dom}$, $\xi$ defined as  
\begin{equation}
\label{a2bis17072017}
\xi \equiv \left(\frac{B}{{\rm nG}} \right) \left(g_{a \gamma \gamma} \, 10^{11} \, {\rm GeV} \right)~,
\end{equation}
and the ALP mass $m$. In particular, owing to the previous bounds on $B$ and $g_{a \gamma \gamma}$ Eq. (\ref{a2bis17072017}) yields $\xi < 11.22$. In order to be specific, as benchmark values we take $L_{\rm dom} = 4 \, {\rm Mpc}$, $10 \, {\rm Mpc}$, and in order not to stay too close to the latter bound we choose $\xi = 0.1, 0.5, 1, 5$. Note that the advantage to deal with $\xi$ is that we do not have to commit ourselves with the values of $B$ and $g_{a \gamma \gamma}$ separately (the above preferred range for $B$ is considerably uncertain and nothing is known about the actual value of $g_{a \gamma \gamma}$).
  
Finally, the photon-ALP conversion probability is maximal -- and energy-independent as well as $m$-independent -- provided that we are in the {\it strong mixing} regime: as we carefully discuss in Appendix 
2 of the SM this requirement translates into the constraint $m = {\cal O} (10^{- 10} \, {\rm eV})$. 

We emphasize that for the considered strength of ${\bm B}$ no cascade induced by $e^+$, $e^-$ -- produced in the process $\gamma + \gamma \to e^+ + e^-$ -- is expected to be seen around a source or along the line of sight, indeed in agreement with observations. We also show in Appendix 2 of the SM that ALPs do not interact with the ionized intergalactic medium.

\bigskip 

\centerline{ \ \ \ * \ \ \ * \ \ \ * \ \ \ }

\medskip

At this point -- knowing $P_{\gamma \to \gamma}^{\rm ALP} (E_0, z)$ from Appendix 2 of the SM -- we proceed as in Sect. 3. To wit, we first rewrite Eq. (\ref{a1}) with $P_{\gamma \to \gamma} (E_0, z) \to P_{\gamma \to \gamma}^{\rm ALP} (E_0, z)$ and $\Phi_{\rm em}  \bigl(E_0 (1 + z) \bigr) \to \Phi_{\rm em}^{\rm ALP} \bigl(E_0 (1 + z) \bigr)$, which for our purposes can be recast into the form 
\begin{eqnarray}
\label{a2bis}
&\displaystyle \Phi_{\rm em}^{\rm ALP} \bigl(E_0 (1+z) \bigr) = \Bigl(P_{\gamma \to \gamma}^{{\rm ALP}} (E_0, z) \Bigr)^{- 1}\times \\ 
&\displaystyle \times K_{\rm obs} (z) \left(\frac{E_0}{E_{0,*}} \right)^{- \Gamma_{\rm obs}(z)}~, \nonumber
\end{eqnarray}
owing to Eq. (\ref{a417072017}). Next -- as in Sect. 3 -- we best-fit $\Phi_{\rm em}^{\rm ALP} \bigl(E_0 (1+z) \bigr)$ to the single power-law expression 
\begin{equation}    
\label{a418q}
\Phi_{\rm em}^{\rm ALP, BF} \bigl(E_0 (1+z) \bigr) = K_{\rm em}^{\rm ALP} (z) \, 
\left(\frac{E_0 (1 + z)}{E_{0,*}} \right)^{- \Gamma_{\rm em}^{\rm ALP} (z)}
\end{equation}
over the energy range $\Delta E_0 (z)$ where a source is observed, hence $E_0$ varies within $\Delta E_0 (z)$. Such a best-fitting procedure is performed for every benchmark value of $\xi$ and $L_{\rm dom}$ specified above. The resulting values of $\Gamma_{\rm em}^{\rm ALP}$ for the blazars in ${\cal S}$ are reported in Table III of the SM for $L_{\rm dom} = 4 \, {\rm Mpc}$, $\xi = 0.1, 0.5, 1, 5$, and in Table IV of the SM for $L_{\rm dom} = 10 \, {\rm Mpc}$,  $\xi = 0.1, 0.5, 1, 5$.

We proceed to carry out a statistical analysis of the values of $\Gamma_{\rm em}^{\rm ALP} (z)$ for all blazars in ${\cal S}$, again for any benchmark value of $\xi$ and $L_{\rm dom}$. We use the least square method already employed in Sect. 3 and we try to fit the data with one parameter (horizontal line), two parameters (first-order polynomial) and three parameters (second-order polynomial). 

Finally, in order to quantify the statistical significance of each fit we compute the corresponding $\chi^2_{\rm red, ALP}$, whose values are reported in Table~\ref{tabChiL4} for $L_{\rm dom} = 4 \, {\rm Mpc}$, $\xi = 0.1, 0.5, 1, 5$, and in Table~\ref{tabChiL10} for $L_{\rm dom} = 10 \, {\rm Mpc}$, $\xi = 0.1, 0.5, 1, 5$. In both Tables the values of $\chi^2_{\rm red, CP}$ are reported for comparison.

\begin{table*}
\begin{center}
\begin{tabular}{l|c|cccc}
\hline
\multicolumn{1}{c|}{\# of fit parameters} &\multicolumn{1}{c|}{$\chi^2_{\rm red, CP}$} &\multicolumn{4}{c}{$\chi^2_{\rm red, ALP}$} \\
\hline
\hline

& & \ \  $\xi=0.1$ \ \ & \ \  $\xi=0.5$ \ \ & \ \  $\xi=1$ \ \ & \ \  $\xi=5$ \ \ \\
1 & 2.37 & 2.29 & {\bf 1.29} & 1.31 & 1.43 \\
2 & 1.49 & 1.47 & 1.29 & 1.31 & 1.38 \\
3 & 1.46 & 1.46 & 1.32 & 1.31 & 1.37 \\

\hline
\end{tabular}
\caption{Values of $\chi^2_{\rm red, CP}$ in the case of conventional physics and $\chi^2_{\rm red, ALP}$ within the ALP scenario, for all blazars belonging to ${\cal S}$. The first column indicates the number of fit parameters, the second column concerns  conventional physics and the third column refers to the ALP scenario for $L_{\rm dom} = 4 \, {\rm Mpc}$ and our benchmark values of $\xi$. The number in boldface corresponds to the minimum of $\chi^2_{\rm red, ALP}$.}
\label{tabChiL4}
\end{center}
\end{table*}


\begin{table*}
\begin{center}
\begin{tabular}{l|c|cccc}
\hline
\multicolumn{1}{c|}{\# of fit parameters} &\multicolumn{1}{c|}{$\chi^2_{\rm red, CP}$} &\multicolumn{4}{c}{$\chi^2_{\rm red, ALP}$} \\
\hline
\hline

& & \ \  $\xi=0.1$ \ \ & \ \  $\xi=0.5$ \ \ & \ \  $\xi=1$ \ \ & \ \  $\xi=5$ \ \ \\
1 & 2.37 & 2.05 & {\bf 1.25} & 1.39 & 1.43 \\
2 & 1.49 & 1.44 & 1.26 & 1.37 & 1.38 \\
3 & 1.46 & 1.46 & 1.28 & 1.36 & 1.37 \\

\hline
\end{tabular}
\caption{Same as Table~\ref{tabChiL4} but for $L_{\rm dom} = 10 \, {\rm Mpc}$.}
\label{tabChiL10}
\end{center}
\end{table*}

\begin{figure*}           
\begin{center}
\includegraphics[width=.50\textwidth]{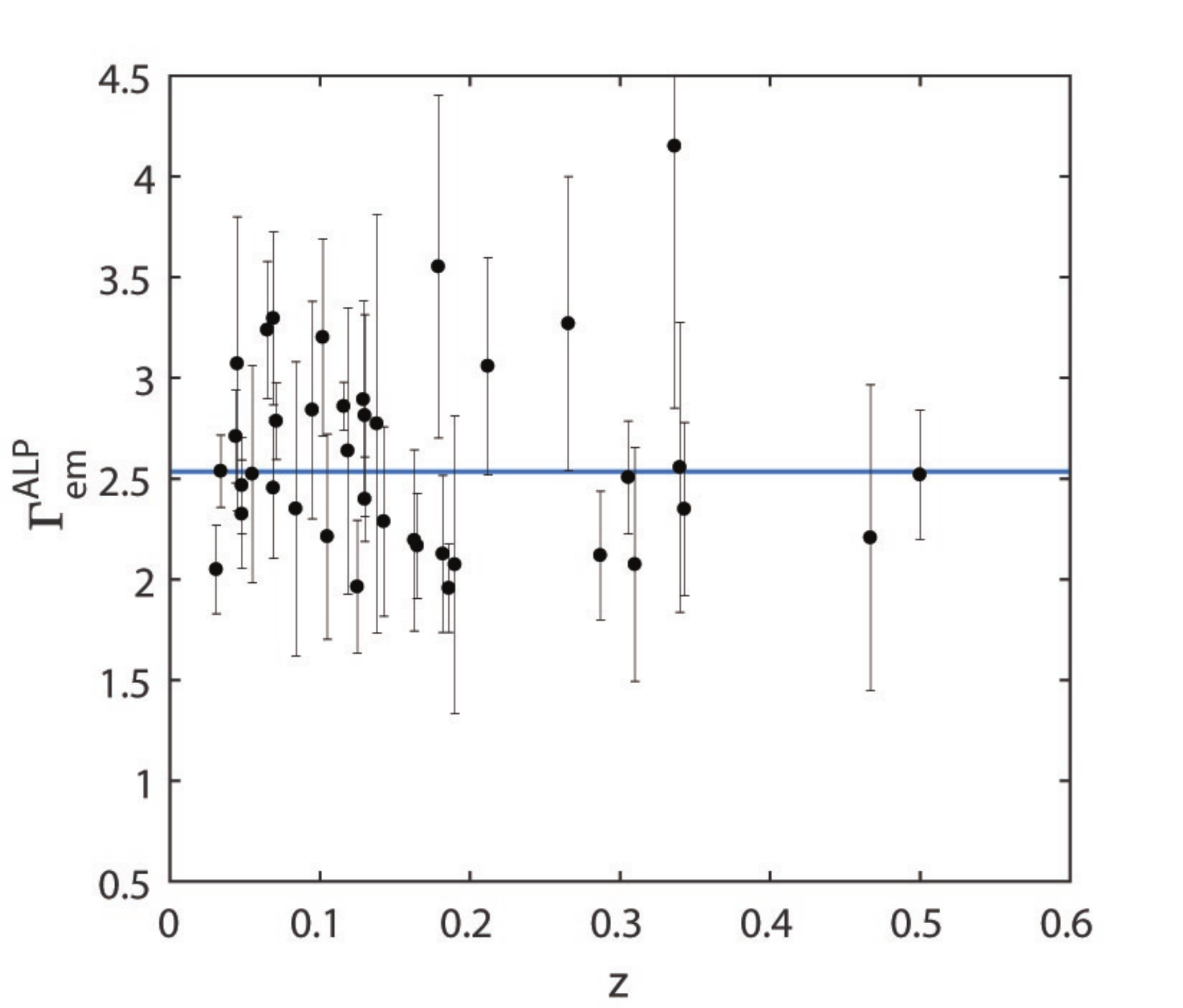}\includegraphics[width=.50\textwidth]{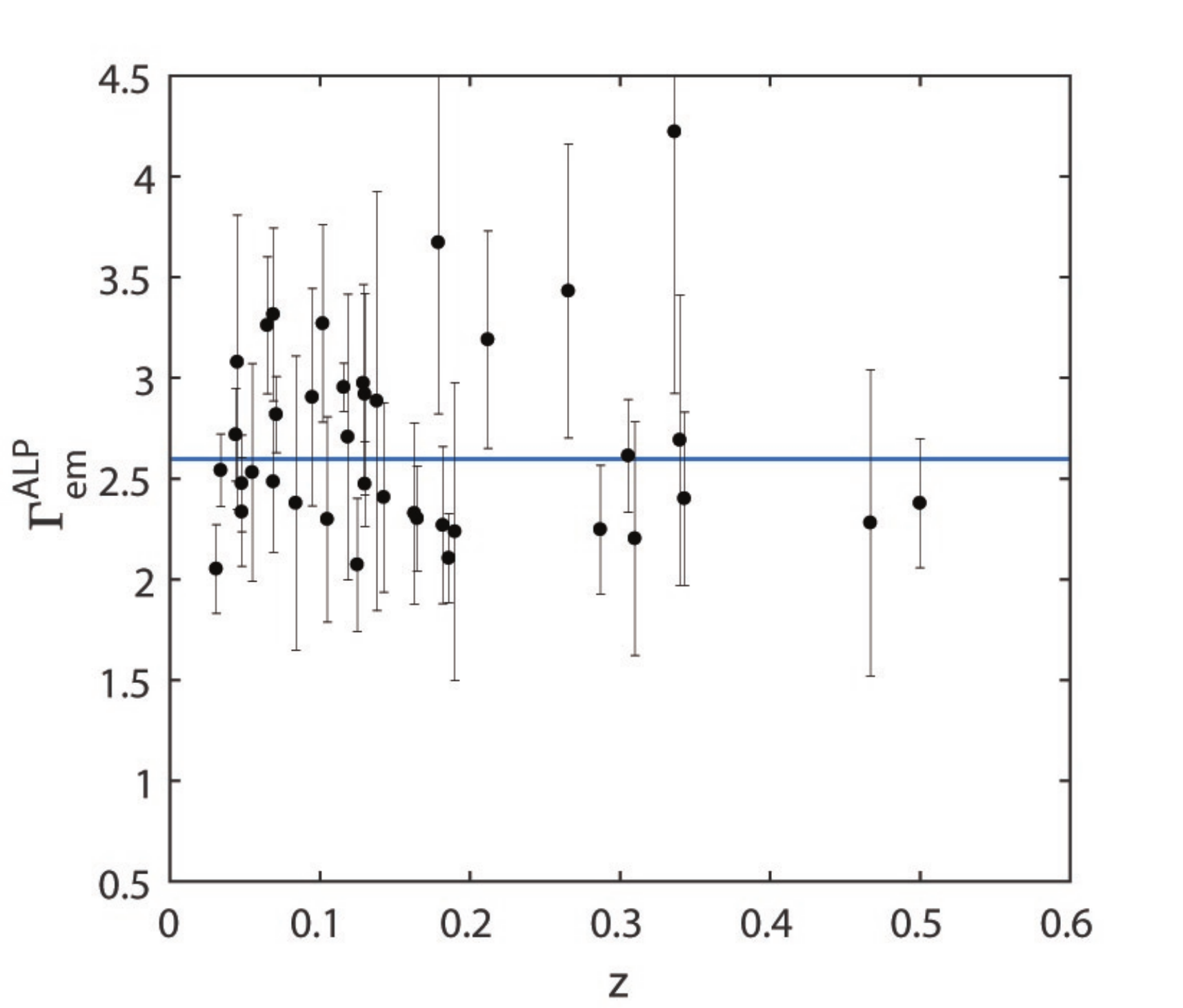}
\end{center} 
\caption{\label{fig5} {\it Left panel}: the values of $\Gamma_{\rm em}^{\rm ALP}$ with the corresponding error bars are plotted versus $z$ for all considered blazars in the case $L_{\rm dom} = 4 \, {\rm Mpc}$, $\xi = 0.5$. Superimposed is the horizontal straight best-fit regression line with $\Gamma_{\rm em}^{\rm ALP} = 2.54$ and $\chi^2_{\rm red,ALP} = 1.29$. {\it Right panel}: Same as left panel, but corresponding to the case $L_{\rm dom} = 10 \, {\rm Mpc}$, $\xi = 0.5$. Superimposed is the horizontal straight best-fit regression line 
with $\Gamma_{\rm em}^{\rm ALP} = 2.60$ and $\chi^2_{\rm red,ALP} = 1.25$.}
\end{figure*}

Basically, such a statistical procedure singles out two preferred situations (corresponding to the minimum of $\chi^2_{\rm red, ALP}$): one for $L_{\rm dom} = 4 \, {\rm Mpc}$ and the other for $L_{\rm dom} = 10 \, {\rm Mpc}$. In either case, we get $\xi = 0.5$ and a {\it straight} best-fit regression line which is exactly {\it horizontal} (in the case of equal $\chi^2_{\rm red, ALP}$ values we obviously choose the model with the lower number of fit parameters). Specifically, for $L_{\rm dom} = 4 \, {\rm Mpc}$ we find $\chi^2_{\rm red, ALP} = 1.29$ and $\Gamma_{\rm em}^{\rm ALP} = 2.54$, while for $L_{\rm dom} = 10 \, {\rm Mpc}$ we obtain $\chi^2_{\rm red, ALP} = 1.25$ and $\Gamma_{\rm em}^{\rm ALP} = 2.60$. Manifestly, both cases turn out to be very similar. We plot the values of $\Gamma_{\rm em}^{\rm ALP} (z)$ in Fig.~\ref{fig5} only for the two considered situations.

Thanks to the fact that $\xi = 0.5$ is our preferred value, we are now in position to make a sharp prediction of the ALP parameters. As it is shown in Appendix 2 of the SM the ALP mass must be $m = {\cal O} 
(10^{- 10} \, {\rm eV})$. Moreover, by employing Eq. (\ref{a2bis17072017}) with $\xi = 0.5$ and recalling the upper bounds on $g_{a \gamma \gamma}$ and $B$ we get $2.94 \times 10^{- 12} \, {\rm GeV}^{- 1} < g_{a \gamma \gamma} < 0.66 \times 10^{- 10} \, {\rm GeV}^{- 1}$. 

\begin{figure*}
\begin{center}
\includegraphics[width=.50\textwidth]{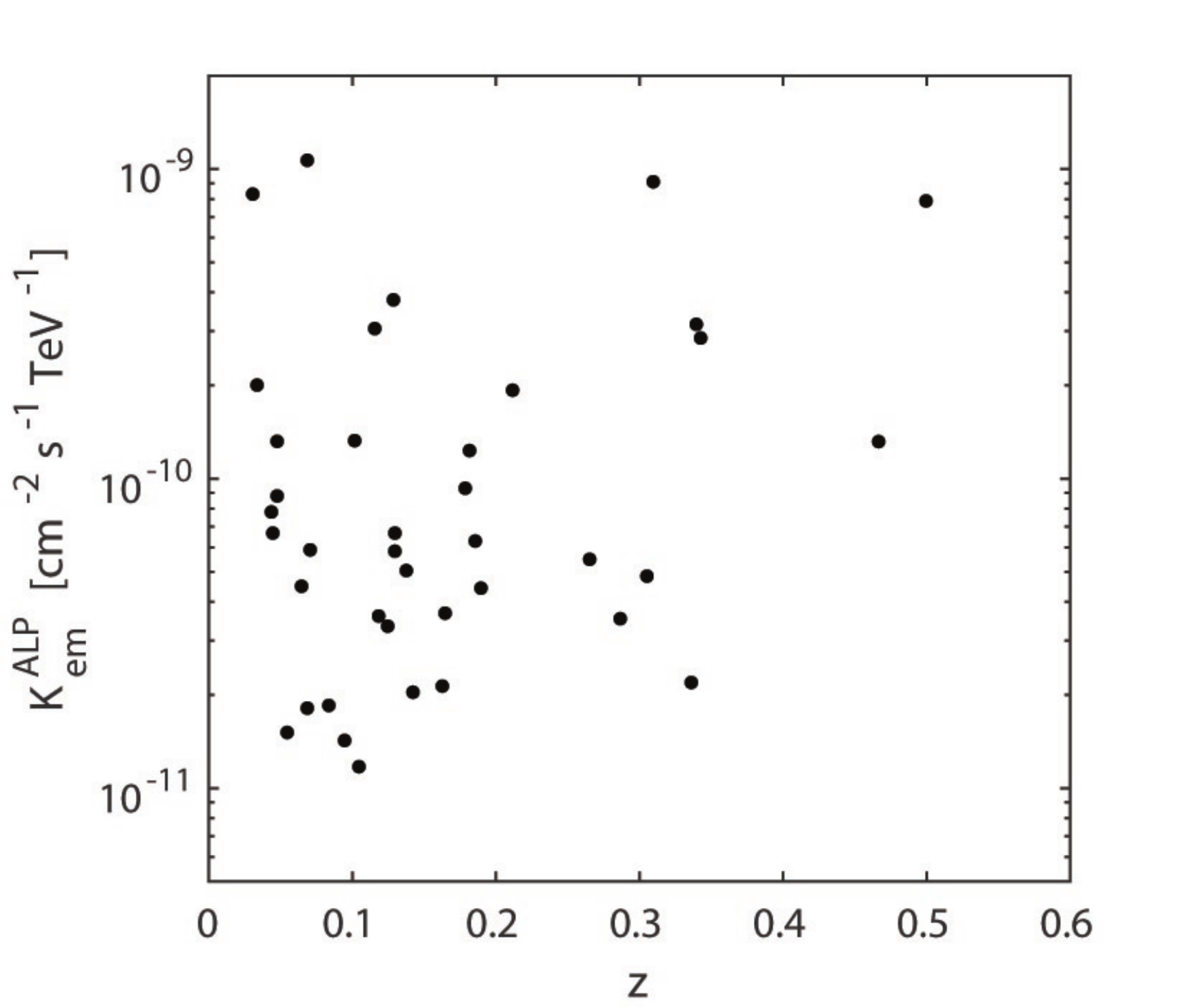}\includegraphics[width=.50\textwidth]{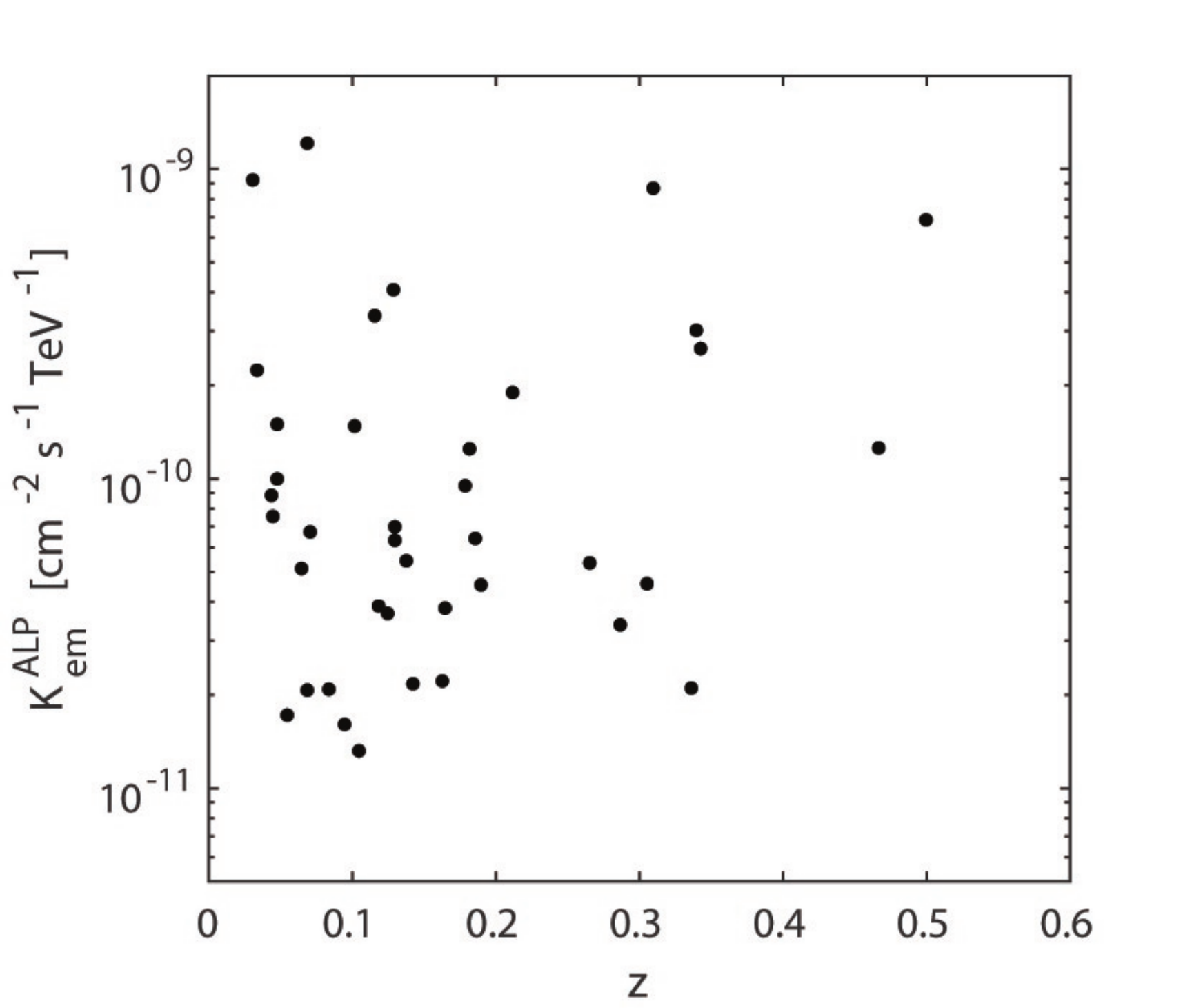}
\end{center} 
\caption{\label{fig2SM} The values of $K_{\rm em}^{\rm ALP}$ are plotted versus $z$ for all considered blazars belonging to $\cal S$. {\it Left panel}: Case $L_{\rm dom} = 4 \, {\rm Mpc}$ and $\xi = 0.5$. {\it Right panel}: Case $L_{\rm dom} = 10 \, {\rm Mpc}$ and $\xi = 0.5$.}
\end{figure*}

In view of our later needs, we also report the plot the values of $K^{\rm ALP}_{\rm em} (z)$ from Table V of the SM in Fig.~\ref{fig2SM}. 

Yet, before claiming that our goal has been achieved, we have to make sure that the above results do not arise from a selection bias. This is done in Appendix 4 of the SM.

As a consequence, we have indeed succeeded in getting rid of the VHE BL Lac spectral anomaly. We stress that it is an {\it automatic} consequence of the ALP scenario, and not an {\it ad hoc} requirement as in the case discussed in Sect. 5. 

We remark in passing that it can be shown that for any given choice of $L_{\rm dom}$ the above statistical procedure fixes the values of $\xi$ and $\chi^2_{\rm red, ALP}$. Even though we have taken realistic values for $L_{\rm dom}$, in the lack of any reliable information about its actual value it can well happen that other interesting results can emerge for different values of $L_{\rm dom}$.

It is convenient to discuss the redshift dependence of the slope difference $\Gamma_{\rm obs} (z) - \Gamma^{\rm ALP}_{\rm em} (z)$ in the previous case in Appendix 5 of the SM, in order to facilitate the comparison with the case of conventional physics.

\section{A NEW SCENARIO FOR FLARING BL LACS}

Besides getting rid of the VHE BL Lac spectral anomaly, the ALP scenario naturally leads to a new view of flaring BL Lacs.

In order to better appreciate this point, it is enlightening to recall what we found in Sect. 5 within conventional physics. We have fitted the values of $\Gamma_{\rm em}^{\rm CP} (z)$ by a horizontal straight regression line, at the price of relaxing the best-fitting requirement. Accordingly, we have seen that the scatter of the values of $\Gamma_{\rm em}^{\rm CP} (z)$ for $95 \, \%$ of blazars belonging to ${\cal S}$ is less than $20 \, \%$ of the mean value set by horizontal straight regression line, namely equal to 0.47. Superficially, the VHE BL Lac spectral anomaly problem would be solved -- but in reality is not -- since we correspondingly have $\chi^2_{\rm red, CP} = 2.37$ which is by far too large.

\begin{figure*}
\centering
\includegraphics[width=.50\textwidth]{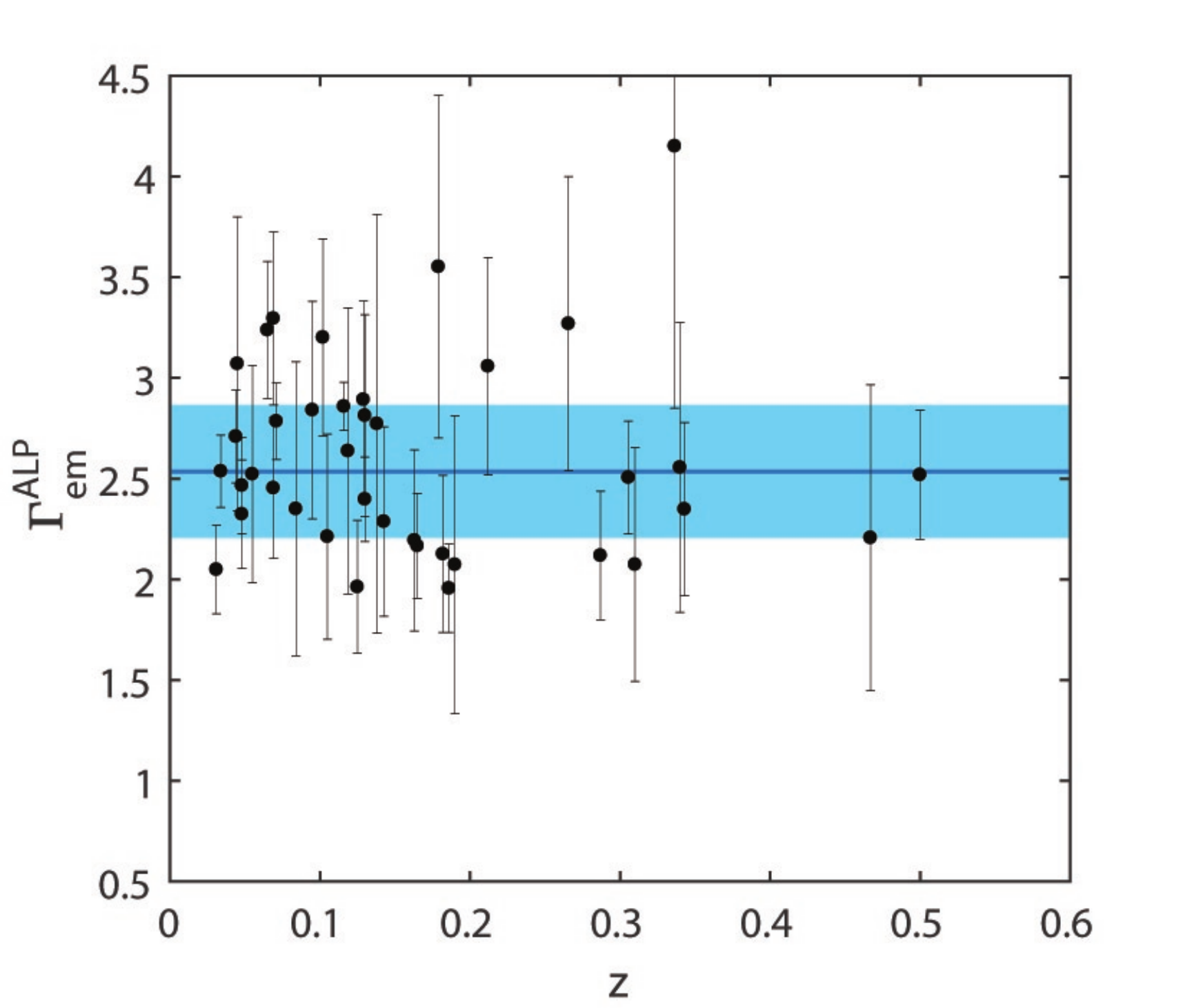}\includegraphics[width=.50\textwidth]{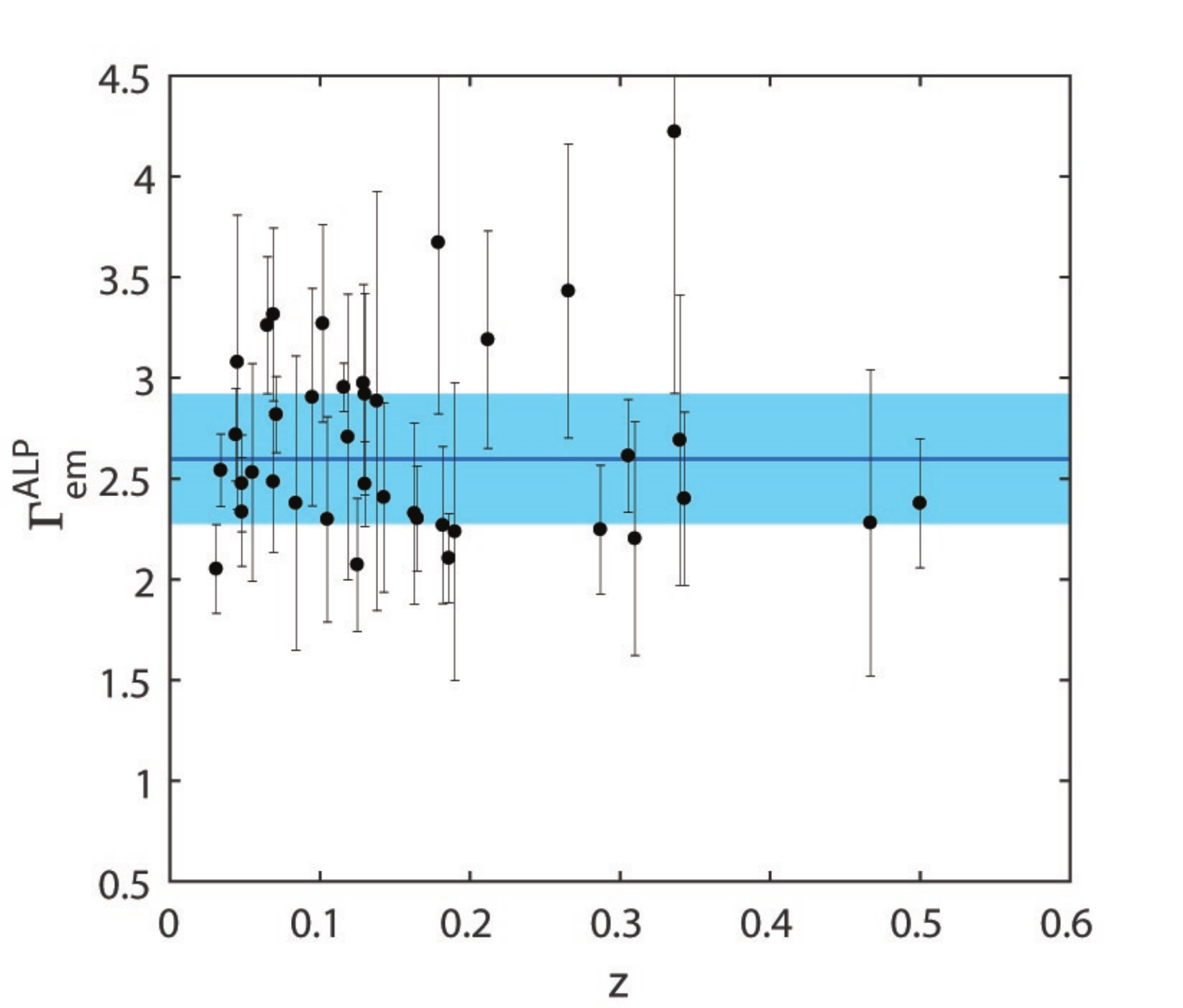}
\caption{\label{fig505} The values of $\Gamma_{\rm em}^{\rm ALP}$ with the corresponding error bars are plotted versus $z$ for all considered blazars within $\cal S$ in the ALP scenario. The light blue strip encompasses $95 \, \%$ of the sources. {\it Left panel}: Case $L_{\rm dom} = 4 \, {\rm Mpc}$. Superimposed is the horizontal straight best-fit regression line $\Gamma_{\rm em}^{\rm ALP} = 2.54$ with $\chi^2_{\rm red, ALP} = 1.29$ and the width of the light blue strip is $\Delta \Gamma_{\rm em}^{\rm ALP} = 
0.66$ which equals  $26 \, \%$ of the value $\Gamma_{\rm em}^{\rm ALP} = 2.54$. {\it Right panel}: Case $L_{\rm dom} = 10 \, {\rm Mpc}$. Superimposed is the horizontal straight best-fit regression line $\Gamma_{\rm em}^{\rm ALP} = 2.60$ with $\chi^2_{\rm red, ALP} = 1.25$ and the width of the light blue strip is $\Delta \Gamma_{\rm em}^{\rm ALP} = 
0.65$ which equals  $25 \, \%$ of the value $\Gamma_{\rm em}^{\rm ALP} = 2.60$.}
\end{figure*}

The results obtained in Sect. 6 leads to a similar but much more satisfactory picture. In the first place, we are dealing with a horizontal straight {\it best-fit} regression line, and in addition the corresponding $\chi^2_{\rm red, ALP}$ turns out to be considerably smaller. Specifically, the scatter of the values of $\Gamma_{\rm em}^{\rm ALP} (z)$ for $95 \, \%$ of the considered blazars is now less than $13 \, \%$ about the mean value set by $\Gamma_{\rm em}^{\rm ALP} = 2.54$ for $L_{\rm dom} = 4 \, {\rm Mpc}$ and less than $13 \, \%$ about the mean value set by $\Gamma_{\rm em}^{\rm ALP} = 2.60$ for $L_{\rm dom} = 10 \, {\rm Mpc}$, namely equal to 0.33 for $L_{\rm dom} = 4 \, {\rm Mpc}$ and equal to 0.32 for $L_{\rm dom} = 10 \, {\rm Mpc}$. This situation is shown in Fig.~\ref{fig505}.

We argue that the small scatter in the values of $\Gamma_{\rm em}^{\rm ALP} (z)$ implies that the physical emission mechanism is the same for all flaring blazars, with the small fluctuations in $\Gamma_{\rm em}^{\rm ALP} (z)$ arising from the difference of their {\it internal} quantities: after all, no two identical galaxies have ever been found! On the other hand, the larger scatter in the values of $K^{\rm ALP}_{\rm em} (z)$ shown in Fig.~\ref{fig2SM} -- presumably unaffected by photon-ALP oscillations when error bars are taken into account -- is naturally traced back to the different environmental state of each flaring source, such as for instance the accretion rate. 

A natural question finally arises. How is it possible that the large spread in the $\{\Gamma_{\rm obs} (z) \}$ distribution exhibited in Fig.~1 arises from the small scatter in the $\{\Gamma_{\rm em}^{\rm ALP} (z) \}$ distribution shown in Fig.~\ref{fig5}? The answer is very simple: most of the scatter in the $\{\Gamma_{\rm obs} (z) \}$ distribution arises from the large scatter in the source redshifts.

\section{CONCLUSIONS}

We have performed a preliminary, comparative analysis of the most homogeneous set of all IBL and HBL flaring BL Lacs for which the redshift, the observed spectrum and the energy range wherein they are observed are known. In order to avoid cosmological evolutionary effects in the considered blazars, we have restricted our sample ${\cal S}$ to the redshift range $0 \leq z \leq 0.6$. We have addressed a possible correlation between their VHE emitted spectra and their redshift. Our analysis has been carried out within the standard photon emission models.       

Given the exploratory nature of our investigation, we have made two simplifying assumptions: we have assumed that all observed VHE spectra of the blazars in ${\cal S}$  can be fitted by a single power-law (neglecting a possible small curvature of some spectra in their lowest energy part, see Eq. (\ref{a417072017})), and we have inferred the emitted spectra by deabsorbing the observed ones and best-fitting the result to a single power law (see Eqs. (\ref{a418}) and  (\ref{a418q})). 

According to a logical standpoint, we have obtained two different results. 

\begin{enumerate}

\item Working within conventional physics (CP), we have discovered a {\it statistical correlation} between the $\{\Gamma_{\rm em}^{\rm CP} (z) \}$ distribution and $z$. We have strongly motivated our belief that such a statistical correlation does not arise from the EBL-absorption. We have also demonstrated that the above correlation is not due to the volume bias. The statistical correlation in question has a best-fit regression line given by a concave parabola decreasing as $z$ increases with $\chi^2_{\rm red, CP} = 1.46$, which implies that blazars with harder spectra are found on average {\it only} at larger redshifts. Therefore, the obvious question arises: how can each source get to know its $z$ and manages to arrange its $\Gamma_{\rm em}^{\rm CP} (z)$ so as to reproduce the above statistical correlation? We are unable to conceive a physical mechanism that is responsible for such a statistical correlation. Hence, we expect the $\{\Gamma_{\rm em}^{\rm CP} (z) \}$ distribution to be statistically uncorrelated with $z$: geometrically, this means that the best-fit regression line of the $\{\Gamma_{\rm em} \}$ distribution should be a {\it straight horizontal line} in the $\Gamma_{\rm em} - z$ plane. We have called the existence of such a correlation the {\it VHE BL Lac spectral anomaly} (which concerns flaring sources alone). We have tried to get rid of it by fitting the data with a horizontal straight regression line $\Gamma_{\rm em}^{\rm CP} = 2.41$. However -- since it does not best-fit the data -- we have found $\chi^2_{\rm red, CP} = 2.37$, which is unacceptably large. Thus, we do not see any way out of such conundrum within conventional physics.

\item As an attempt to get rid of the VHE BL Lac spectral anomaly, we have worked within the ALP scenario, where photon-ALP oscillations in extragalactic space reduce the 
EBL absorption. We have assumed the standard domain-like morphology for the extragalactic magnetic field ${\bm B}$ with coherence length $1 \, {\rm Mpc} \lesssim L_{\rm dom} \lesssim 10 \, {\rm Mpc}$ and the ${\bm B}$ strength in the range $0.1 \, {\rm nG} \lesssim B \lesssim  1 \, {\rm nG}$, as strongly suggested by quasar and primeval galactic outflows models. Moreover, assuming for the ALP mass $m = {\cal O} (10^{- 10} \, {\rm eV})$ and the two-photon coupling in the range $2.94 \times 10^{- 12} \, {\rm GeV}^{- 1} < g_{a \gamma \gamma} < 0.66 \times 10^{- 10} \, {\rm GeV}^{- 1}$ we have shown that the VHE BL Lac spectral anomaly disappears altogether, since the best-fit regression line becomes exactly {\it straight and horizontal} in the $\Gamma_{\rm  em} - z$. Hence, the effect of photon-ALP oscillations in extragalactic space combined with the EBL-absorption is to turn it into the horizontal straight best-fit regression lines $\Gamma_{\rm em}^{\rm ALP} = 2.54$ and $\Gamma_{\rm em}^{\rm ALP} = 2.60$ for $L_{\rm dom} = 4 \, {\rm Mpc}$ and $L_{\rm dom} = 10 \, {\rm Mpc}$, respectively (both cases correspond to $\xi = 0.5$). Beside elegantly solving the VHE BL Lac spectral anomaly problem, this result looks astonishing. Of course, by changing the effective level of EBL absorption we expect the $z$-dependence of the best-fit regression line of the $\{\Gamma_{\rm em}^{\rm ALP} (z)\}$ distribution to differ from that of the $\{\Gamma_{\rm em}^{\rm CP} (z) \}$ distribution. As a consequence, the shape of the best-fit straight regression line in the $\Gamma_{\rm em} - z$ plane changes. But to become exactly straight and horizontal -- which is the {\it only possibility} in agreement with our expectation out of infinitely-many ones -- is a highly nontrivial fact! We emphasize that if this result were due to a selection bias, then it would mean that the combined energy sensitivity threshold of H.E.S.S., MAGIC and VERITAS should indeed be exceptional! Actually, even from a purely statistical point of view the ALP scenario is better than the conventional one, in which we have found $\chi^2_{\rm red, CP} = 1.46$ for the best-fit regression line (Sect. 3) or $\chi^2_{\rm red, CP} = 2.37$ for the horizontal straight fitting line (Sect. 5). Instead, with photon-ALP oscillations the resulting horizontal straight best-fit regression lines have $\chi^2_{\rm red, ALP} = 1.29$ for $L_{\rm dom} = 4 \, {\rm Mpc}$ and $\chi^2_{\rm red, ALP} = 1.25$ for $L_{\rm dom} = 10 \, {\rm Mpc}$ (see Sect. 6). Finally, a new scenario for flaring blazars arises, in which it is natural to suppose that the slope $\Gamma_{\rm em}^{\rm ALP} (z)$ is governed by fundamental physics and the internal BL Lac properties, while the normalization constant $K_{\rm em}^{\rm ALP} (z)$ depends on the environmental conditions. 

\end{enumerate}

Moreover, it turns out that for the same values of the model parameters which lead to the above conclusion ALPs also explain a very different result. According to conventional models, flat spectrum radio quasars (FSRQs) should not emit above $30 \, {\rm GeV}$ at most. This is due to the fact that higher energy photons accelerated in the jet enter -- at a distance of about $10^{18} \, {\rm cm}$ from the centre -- the so-called broad-line region (BLR), whose high density of ultraviolet photons gives rise to an optical depth $\tau \simeq 15$ owing to the same $\gamma + \gamma \to e^+ + e^-$ absorption process which is responsible for the EBL-absorption in extragalactic space. Yet, several FSRQs have been detected at energies up to $400 \, {\rm GeV}$, and so this fact poses a serious challenge for conventional models. A striking case concerns the FSRQ PKS 1222 + 216, which has been observed simultaneously by Fermi/LAT in the energy range $0.3 \, {\rm GeV} \lesssim E \lesssim 3 \, {\rm GeV}$ and by MAGIC in the energy range $70 \, {\rm GeV} \lesssim E \lesssim 400 \, {\rm GeV}$. Actually, MAGIC has detected an intense VHE emission that doubles in only about 10 minutes, thereby implying that the size of the emitting region is about $10^{14} \, {\rm cm}$: apparently such a small blob very far from the centre emits like a whole BL Lac! So far, only {\it ad hoc} astrophysical explanations have been put forward, in the sense that they are devised {\it only} to explain such an effect~\citep{tavecchioetal2011,etropoulou2017}. Instead, a natural explanation emerges within conventional blazar photon emission models containing ALPs, since photon-ALP oscillations substantially lower the photon absorption inside the BLR, thereby allowing VHE photons to escape from the BLR and be emitted, in remarkable quantitative agreement with observations~\citep{trgb2012}.

Thus, the combination of the two very different results -- taken at face value -- provides a 
hint at an ALP with mass $m = {\cal O} (10^{-10} \, {\rm eV})$ and two-photon 
coupling in the range $2.94 \times 10^{- 12} \, {\rm eV} < g_{a \gamma \gamma} < 0.66 \times 10^{- 10} \, {\rm eV}$, which is consistent with all available constraints (as shown in Appendix 2 of the SM). And last but not least, the Universe correspondingly becomes considerably more transparent above energies $E \gtrsim 1 \, {\rm TeV}$ than as implied by conventional physics.

We would like to point out that our proposal is quite predictive and can be checked not only with the advent of new gamma-ray detectors like the CTA (Cherenkov Telescope Array)\footnote{https://www.cta-observatory.org/}, HAWC (High-Altitude Water Cherenkov Observatory)\footnote{http://www.hawc-observatory.org/}, GAMMA-400 (Gamma Astronomical Multifunctional Modular Apparatus)\footnote{gamma400.lebedev.ru/gamma$400_{\rm e}$.html}, LHAASO (Large High Altitude Air Shower Observatory)\footnote{http://english.ihep.cas.cn/ic/ip/LHAASO/}, TAIGA-HiSCORE (Hundred Square km Cosmic Origin Explorer)\footnote{www.desy.de/groups/astroparticle/score/en/} and HERD (High Energy Cosmic Radiation Detection,~\citealt{herd}), but also thanks to laboratory data based on the planned experiments ALPS II (Any Light Particle Search, upgraded)\footnote{https://alps.desy.de/e141064}, STAX (Sub-THz-AXion,~\citealt{stax}) and IAXO (International Axion Observatory,~\citealt{iaxo,iaxo1}), which will have the capability to discover or rule out an ALP with the properties assumed in the present analysis. The same is true for laboratory experiments exploiting the techniques discussed in~\citet{avignone1,avignone2,avignone3}.     

We again emphasize that the whole analysis reported in this paper should be regarded as a preliminary attempt, and a more thorough discussion is needed to assess our conclusions. 

Nevertheless, our Universe may in this way be offering us a compelling reason to push physics beyond the Standard Model along a very specific direction, which possibly sheds light also on the nature of cold dark matter.

\section*{ACKNOWLEDGMENTS}

We thank Fabrizio Tavecchio for comments and Marco Tavani for a careful reading of the manuscript. We want to express our deep gratitude to Giancarlo Setti for having shared with us his efforts on the interpretation of the origin of the VHE BL Lac spectral anomaly. GG acknowledges contribution from the grant INAF CTA--SKA, ``Probing particle acceleration and $\gamma$-ray propagation with CTA and its precursors" and the INAF Main Stream project ``High-energy extragalactic astrophysics: toward the Cherenkov Telescope Array'', while the work of MR is supported by an INFN TAsP grant.

\label{lastpage}

\end{document}